\documentstyle[12pt,epsfig]{article}
\def\be{\begin{equation}}
\def\ee{\end{equation}}
\def\bc{\begin{center}}
\def\ec{\end{center}}
\def\bea{\begin{eqnarray}}
\def\eea{\end{eqnarray}}
\def\dd{\displaystyle}
\def\nn{\nonumber}

\def\gev{{\rm \; GeV}}
\def\h1b{{\overline{H}_1}}
\def\h2b{{\overline{H}_2}}

\def\l1{{\Lambda_1}}

\def\ov{\overline}
\def\simlt{\stackrel{<}{{}_\sim}}
\def\simgt{\stackrel{>}{{}_\sim}}

\def\tev{{\rm \; TeV}}

\def    \bc            {\begin{center}}
\def    \ec            {\end{center}}

\def    \dd            {\displaystyle}
\def    \be            {\begin{equation}}
\def    \ee            {\end{equation}}
\def    \bea           {\begin{eqnarray}}
\def    \eea           {\end{eqnarray}}
\def    \nn            {\nonumber}

\def    \M2            {\mbox{$ \tilde{\cal M} $}}

\def\ov{\overline}
\def\gev{{\rm \; GeV}}
\def\tev{{\rm \; TeV}}

\def\simlt{\stackrel{<}{{}_\sim}}
\def\simgt{\stackrel{>}{{}_\sim}}

\def\aac{{\left[{v_2 c_\beta \over v_R N_3}-{(v_2 a-v_1)s_\beta
\over v_R N_4}\right]}}
\def\aad{{\left[{v_2 s_\beta \over v_R N_3}+{(v_2 a-v_1)c_\beta
\over v_R N_4}\right]}}
\def\abc{{\left[{-v_1 c_\beta \over v_L N_3}+{(v_1 a-v_2)s_\beta
\over v_L N_4}\right]}}
\def\abd{{\left[{-v_1 s_\beta \over v_L N_3}-{(v_1 a-v_2)c_\beta
\over v_L N_4}\right]}}
\def\acc{{\left[{c_\beta \over N_3}-{ a s_\beta
\over N_4}\right]}}
\def\acd{{\left[{s_\beta \over N_3}+{ a c_\beta
\over N_4}\right]}}
\newcommand{\nsect}{\setcounter{equation}{0}
\def\theequation{\thesection.\arabic{equation}}\section}

\newcommand{\appendixA}{\setcounter{equation}{0}
\def\theequation{\rm{A}.\arabic{equation}}\section*}
\newcommand{\appendixB}{\setcounter{equation}{0}
\def\theequation{\rm{B}.\arabic{equation}}\section*}

\catcode`@=11
\def\marginnote#1{}
\newcount\hour
\newcount\minute
\newtoks\amorpm
\hour=\time\divide\hour by60
\minute=\time{\multiply\hour by60 \global\advance\minute by-\hour}
\edef\standardtime{{\ifnum\hour<12 \global\amorpm={am}%
        \else\global\amorpm={pm}\advance\hour by-12 \fi
        \ifnum\hour=0 \hour=12 \fi
        \number\hour:\ifnum\minute<10 0\fi\number\minute\the\amorpm}}
\edef\militarytime{\number\hour:\ifnum\minute<10 0\fi\number\minute}
\def\draftlabel#1{{\@bsphack\if@filesw {\let\thepage\relax
   \xdef\@gtempa{\write\@auxout{\string
      \newlabel{#1}{{\@currentlabel}{\thepage}}}}}\@gtempa
   \if@nobreak \ifvmode\nobreak\fi\fi\fi\@esphack}
        \gdef\@eqnlabel{#1}}
\def\@eqnlabel{}
\def\@vacuum{}
\def\draftmarginnote#1{\marginpar{\raggedright\scriptsize\tt#1}}
\def\draft{\oddsidemargin 0.0truein
        \def\@oddfoot{\sl preliminary draft \hfil
        \rm\thepage\hfil\sl\today\quad\militarytime}
        \let\@evenfoot\@oddfoot \overfullrule 3pt
        \let\label=\draftlabel
        \let\marginnote=\draftmarginnote
   \def\@eqnnum{(\theequation)\rlap{\kern\marginparsep\tt\@eqnlabel}%
\global\let\@eqnlabel\@vacuum}  }
\catcode`@=12
\begin{document}
\begin{titlepage}
\vspace*{-1cm}
hep-ph/9705450
\hfill{DFPD~97/TH/23}
\\
\vskip 1.0cm
\begin{center}
{\Large\bf Phenomenological aspects of a fermiophobic 
\\
\vskip .2cm
$SU(2) \times SU(2) \times U(1)$ extension of the Standard Model}
\end{center}
\vskip 0.8  cm
\begin{center}
{\large Andrea
Donini}\footnote{e-mail address: donini@vxrm70.roma1.infn.it}
\\
\vskip .1cm
Dipartimento di Fisica, Universit\`a `La Sapienza', I-00185 
Rome, Italy
\\
\vskip .5cm
{\large Ferruccio
Feruglio}\footnote{e-mail address: feruglio@padova.infn.it},
{\large Joaquim Matias}\footnote{e-mail address: matias@padova.infn.it}
\\
\vskip .1cm
Dipartimento di Fisica, Universit\`a di Padova, I-35131 Padua, Italy
\\
\vskip .2cm
and
\\
\vskip .2cm
{\large Fabio
Zwirner}\footnote{e-mail address: zwirner@padova.infn.it}
\\
\vskip .1cm
INFN, Sezione di Padova, I-35131 Padua, Italy
\end{center}
\vskip 0.5cm
\begin{abstract}
\noindent
We consider an extension of the standard electroweak theory
with gauge group $SU(2)_L \times SU(2)_R \times U(1)_{\tilde{Y}}$,
where the gauge bosons of the extra $SU(2)_R$ factor do not 
couple to ordinary fermions. We show that precision electroweak 
data and flavour physics provide quite stringent indirect 
constraints on its parameter space, but still allow for 
relatively light non-standard gauge and Higgs bosons. We then 
consider the model phenomenology at high-energy colliders, and 
observe that in the gauge boson sector present bounds and possible 
future signals are dominated by $Z'$ production. In summary, 
indirect constraints on the charged gauge boson sector are so 
tight that observable new effects must be connected either with 
the neutral gauge boson sector or with the extended Higgs sector 
of the model.
\end{abstract}
\vfill{
May 1997}
\end{titlepage}
\setcounter{footnote}{0}
\vskip2truecm
\nsect{Introduction}

Extensions of the Standard Model (SM) of electroweak interactions
based on the gauge group $SU(2)_L \times SU(2)_R \times 
U(1)_{\tilde{Y}}$ (for the time being, the labelling of the 
different factors is purely conventional) have been widely 
discussed in the literature [1--5], with various motivations. 
In particular, these models are a natural framework to parametrize 
the possible existence of additional $W'$ and $Z'$ bosons, 
detectable at present and future colliders.

To limit the number of possibilities, we restrict our attention to 
models with the following properties: (i)~they are non-supersymmetric; 
(ii)~their fermionic sector consists only of $SU(2)$ singlets and 
doublets; (iii)~their Higgs sector consists only of $SU(2)$
singlets, doublets and triplets; (iv)~they admit the standard 
embedding of the electric charge:
\be
\label{charge}
Q = T_{3L} + T_{3R} + \tilde{Y} \, ;
\ee
(v)~the gauge interactions are universal for the three fermion
generations. Even under the above assumptions, a considerable 
freedom remains, which allows for at least five different 
models\footnote{Of course, further models can be constructed by 
relaxing one or more of the previous assumptions: an example is 
the `topflavor' model \cite{tf}, where $SU(2)_L$ acts on the 
first two generations and $SU(2)_R$ on the third one.}:
\begin{itemize}
\item
the `standard' \cite{st} left-right symmetric model (LR);
\item
the `leptophobic' model (LP);
\item
the `hadrophobic' model (HP);
\item
the `fermiophobic' \cite{fp} model (FP);
\item
the `ununified' \cite{un} model (UN).
\end{itemize}
The various models are defined by the transformation
properties of their fermion content with respect to
the gauge group, summarized in table~1.
\begin{table}[htb]
$$
\begin{array}{|c|c|c|c|c|c|}
\hline
& & & & & \\
Field/Model & LR & LP & HP & FP & UN \\
& & & & & \\
\hline
& & & & & \\
q_L \equiv  \left( \begin{array}{c} u_L \\ d_L \end{array} \right)
& (2,1,1/6) & (2,1,1/6) & (2,1,1/6) & (2,1,1/6) & (2,1,1/6) \\
& & & & & \\
\hline
& & & & & \\
q_R \equiv  \left( \begin{array}{c} u_R \\ d_R \end{array} \right)
& (1,2,1/6) & (1,2,1/6) & \begin{array}{c} (1,1,2/3) \\ (1,1,-1/3)
\end{array} & \begin{array}{c} (1,1,2/3) \\ (1,1,-1/3) \end{array}
& \begin{array}{c} (1,1,2/3) \\ (1,1,-1/3) \end{array}  \\
& & & & & \\
\hline
& & & & & \\
l_L \equiv  \left( \begin{array}{c} \nu_L \\ e_L \end{array} \right)
& (2,1,-1/2) & (2,1,-1/2) & (2,1,-1/2) & (2,1,-1/2) & (1,2,-1/2) \\
& & & & & \\
\hline
& & & & & \\
l_R \equiv  \left( \begin{array}{c} \nu_R \\ e_R \end{array} \right)
& (1,2,-1/2) & \begin{array}{c} - \\ (1,1,-1) \end{array} &
(1,2,-1/2)
& \begin{array}{c} - \\ (1,1,-1) \end{array} & \begin{array}{c} - \\
(1,1,-1) \end{array} \\
& & & & & \\
\hline
\end{array}
$$
\caption{{\it Fermion transformation properties in the models 
considered in the text. The numbers in brackets refer to $SU(2)_L$,
$SU(2)_R$ and $U(1)_{\tilde{Y}}$, respectively. Colour and 
generation indices are implicit.}}
\end{table}
Notice that some of the models (LR, HP) include right-handed
neutrinos $\nu_R$, whilst some others (LP, FP, UN) have exactly 
the fermion content of the SM. The Higgs fields that can play a 
role in the spontaneous breaking of the gauge symmetry, assumed 
to proceed according to the following pattern
\be
SU(2)_L \times SU(2)_R \times U(1) \longrightarrow U(1)_{e.m.} \, ,
\ee
are those transforming non-trivially under the gauge group but 
containing at least one electrically neutral component, and are 
listed in table~2. Those needed to get an acceptable tree-level 
fermion mass spectrum are marked with the symbol $\otimes$. 
Others, as we shall see, may be needed to get an acceptable mass 
spectrum in the  gauge boson sector: standard choices are marked 
with the symbol $\times$.
\begin{table}[htb]
$$
\begin{array}{|c|c|c|c|c|c|}
\hline
& & & & & \\
Field/Model & LR & LP & HP & FP & UN \\
& & & & & \\
\hline
& & & & & \\
\phi_{LR} \equiv \left( \begin{array}{cc} \phi_1^0 & \phi_2^+
\\ \phi_1^- & \phi_2^0 \end{array} \right) \sim (2,2,0) 
& \otimes & \otimes & \otimes & \times & \times \\
& & & & & \\
\hline
& & & & & \\
\phi_L \equiv \left( \begin{array}{c} \phi_L^0 \\ \phi_L^-
\end{array} \right) \sim (2,1,-1/2)  & - & \otimes & 
\otimes & \otimes & \otimes \\
& & & & & \\
\hline
& & & & & \\
\phi_R \equiv \left( \begin{array}{c} \phi_R^0 \\ \phi_R^-
\end{array} \right) \sim (1,2,-1/2) 
& - & - & - & \times & \otimes \\
& & & & & \\
\hline
& & & & & \\
\Delta_L \equiv \left( \begin{array}{cc} \frac{1}{\sqrt{2}}
\delta_L^+ & \delta_L^{++} \\ \delta_L^0 & - \frac{1}{\sqrt{2}}
\delta_L^+ \end{array} \right) \sim (3,1,1) 
& \times & - & - & - & - \\
& & & & & \\
\hline
& & & & & \\
\Delta_R \equiv \left( \begin{array}{cc} \frac{1}{\sqrt{2}}
\delta_R^+ & \delta_R^{++} \\ \delta_R^0 & - \frac{1}{\sqrt{2}}
\delta_R^+ \end{array} \right) \sim (1,3,1) 
& \otimes & - & \otimes & - & - \\
& & & & & \\
\hline
\end{array}
$$
\caption{{\it Typical sets of Higgs fields for the models considered
in the text.}}
\end{table}

In the class of models considered above, we would like to select
a candidate model that can naturally satisfy all the existing 
phenomenological constraints and, at the same time, allow for
relatively light extra gauge bosons, accessible to future
accelerators such as the upgraded Tevatron collider and the LHC.
In our opinion, a palatable candidate is the FP model: it is
automatically free of gauge anomalies (in contrast with the LP,
HP and UN models); it does not contain right-handed neutrinos, so
it can do without Higgs triplets and still provide an acceptable 
tree-level fermion and gauge boson mass spectrum (in contrast with 
the LR and HP models); it automatically guarantees the absence of 
flavour-changing neutral currents (FCNC) at tree level, and the 
suppression of loop-induced effects, thanks to the fact that the 
unmixed $SU(2)_R$ gauge bosons and the $(\phi_{LR},\phi_R)$ Higgs 
bosons cannot have gauge-invariant couplings to the matter fermions 
(in contrast with all the other models of our list). 

For the above reasons, in the rest of this paper we restrict our 
attention to the FP model, and present a phenomenological analysis 
of it as complete as possible. In section~2 we discuss the general
structure of the FP model, considering first masses and mixings in 
the various sectors, and then gauge and Yukawa interactions in the
mass eigenstate basis. Section~3 deals with the many facets of the 
FP-model phenomenology: constraints from precision electroweak data
and from flavour physics, as well as production and decay of $W'$ 
and $Z'$ bosons at hadron colliders. Some useful formulae are 
collected in the appendices.

\nsect{General structure of the fermiophobic model}

The FP model is described by a gauge-invariant Lagrangian 
density of the form
\be
{\cal L}={\cal L}_{YM}+{\cal L}_S+{\cal L}_F+{\cal L}_Y \, .
\label{lag}
\ee
The Yang-Mills term ${\cal L}_{YM}$ is given by:
\be
{\cal L}_{YM}=
-\dd\frac{1}{4} {F_L^a}_{\mu\nu}{F_L^a}^{\mu\nu}
-\dd\frac{1}{4} {F_R^a}_{\mu\nu}{F_R^a}^{\mu\nu}
-\dd\frac{1}{4} {B}_{\mu\nu}{B}^{\mu\nu}+ \ldots
\, ,
\label{ym}
\ee
where the dots stand for terms involving the gluons and
\bea
{F_L^a}_{\mu\nu}&=& \partial_{\mu} W_{L \, \nu}^a -
\partial_{\nu} W_{L \, \mu}^a + g_L \epsilon^{abc}
W_{L \, \mu}^b W_{L \, \nu}^c \, , \nn \\
{F_R^a}_{\mu\nu}&=& \partial_{\mu} W_{R \, \nu}^a -
\partial_{\nu} W_{R \, \mu}^a + g_R \epsilon^{abc}
W_{R \, \mu}^b W_{R \, \nu}^c \, , \nn \\
{B}_{\mu\nu}&=&\partial_\mu B_\nu-\partial_\nu B_\mu \, .
\eea
The term ${\cal L}_S$, containing generalized kinetic terms
and self-interactions of the spin--0 fields, is given by:
\be
{\cal L}_S=
(D^\mu \phi_L)^\dagger D_\mu \phi_L+
(D^\mu \phi_R)^\dagger D_\mu \phi_R+
{\rm tr \,} 
\left[(D^\mu \phi_{LR})^\dagger D_\mu \phi_{LR}\right]-V_0 \, ,
\label{ls}
\ee
where $V_0$ is the scalar potential, a gauge-invariant polynomial 
of degree four in the fields ($\phi_L,\phi_R,\phi_{LR}$) and their
hermitean conjugates, and the covariant derivatives read:
\bea
D_\mu \phi_L&=&(\partial_\mu
-i g_L {W^a_L}_\mu\dd\frac{\tau^a}{2}
+ {i \over 2} \tilde{g} B_\mu
)\phi_L \, , \nn\\
D_\mu \phi_R&=&(\partial_\mu
-i g_R {W^a_R}_\mu\dd\frac{\tau^a}{2}
+ {i \over 2} \tilde{g} B_\mu
)\phi_R \, , \nn\\
D_\mu \phi_{LR}&=&\partial_\mu\phi_{LR}
-i g_L {W^a_L}_\mu\dd\frac{\tau^a}{2}\phi_{LR}
+i g_R {W^a_R}_\mu\phi_{LR}\dd\frac{\tau^a}{2} \, .
\eea
The term ${\cal L}_F$ contains the fermion kinetic terms and gauge 
interactions: since in the FP model all fermions are $SU(2)_R$ 
singlets, their gauge interactions are exactly the same as in the 
SM, when expressed in terms of the gauge vector bosons $({W^a_L}_\mu,
B_\mu)$ and of the corresponding coupling constants $(g_L,{\tilde g}$). 
Finally, the term ${\cal L}_Y$ describes the Yukawa interactions. In
the FP model, the only couplings allowed by gauge invariance are those 
between the fermions and the $SU(2)_L$ doublet $\phi_L$, thus ${\cal L}_Y$
coincides with its SM counterpart. In terms of the quark and lepton 
mass eigenstates, represented by three-dimensional vectors in flavour 
space:
\bea
{\cal L}_Y & = &  
- {\sqrt{2} \over v_L} \left[ \left( 
\ov{u_L} M^U_{diag} u_R +
\ov{d_R} M^D_{diag} d_L +
\ov{e_R} M^E_{diag} e_L \right) \phi_L^0
\right. \nn \\
& + & 
\left.
\left( 
\ov{d_L} K^{\dagger} M^U_{diag} u_R 
- \ov{d_R} M^D_{diag} K^{\dagger}  u_L
- \ov{e_R} M^E_{diag} \nu_L \right) \phi_L^- 
+ 
{\rm h.c.} \right] \, ,
\label{yuk}
\eea
where $M^U_{diag} = (m_u, m_c, m_t)$, $M^D_{diag} = (m_d, m_s, 
m_b)$, $M^E_{diag} = (m_e, m_\mu, m_\tau)$, and $K$ is the
Cabibbo-Kobayashi-Maskawa matrix.

\subsection{Mass spectrum}

To discuss the spectrum of the model, we assume that an appropriate
choice of parameters in the scalar potential $V_0$ leads to the 
following pattern of vacuum expectation values (VEVs) for the scalar 
fields:
\be
\label{vevs}
\langle \phi_{LR} \rangle =
{1 \over \sqrt{2}}
\left( \begin{array}{cc} 
v_1 & 0 \\ 0 & v_2
\end{array} \right)
\, ,
\;\;\;\;\;
\langle \phi_L \rangle =
{1 \over \sqrt{2}}
\left( \begin{array}{c} 
v_L \\ 0 \end{array} 
\right) \, ,
\;\;\;\;\;
\langle \phi_R \rangle =
{1 \over \sqrt{2}}
\left( \begin{array}{c} 
v_R \\ 0 \end{array} 
\right) \, .
\ee
We also assume that ($v_L,v_R,v_1,v_2$) are real,
and define the auxiliary quantities:
\be
\label{auxqua}
u^2 \equiv v_1^2 + v_2^2 \, ,
\;\;\;
\tan \beta \equiv {v_2 \over v_1} \, ,
\;\;\;
g \equiv g_L \, ,
\;\;\;
x \equiv {g_R \over g} \, ,
\;\;\;
y \equiv {\tilde{g} \over g} \, .
\ee

Before specializing to the charged and neutral gauge bosons, 
it is convenient to recall the general solution of the 
eigenvalue problem for a $2 \times 2$ mass matrix, in a way 
suitable for taking the physically interesting limit of 
small mixing. In the interaction basis, $(V_1,V_2) \equiv
(V_{SM},V_{extra})$:
\be
{\cal M}^2 =
\left(
\begin{array}{cc}
{\cal M}_{11}
&
{\cal M}_{12}
\\
{\cal M}_{12}
&
{\cal M}_{22}
\end{array}
\right) \, .
\ee
In our conventions, we denote mass eigenvalues and eigenstates by:
\be
m_V^2 = {1 \over 2} \left[ {\cal M}_{11} + {\cal M}_{22}
- \sqrt{ \left( {\cal M}_{11} - {\cal M}_{22} \right)^2 +
4 {\cal M}_{12}^2 } \right] \, ,
\ee
\be
m_{V'}^2 = {1 \over 2} \left[ {\cal M}_{11} + {\cal M}_{22}
+ \sqrt{ \left( {\cal M}_{11} - {\cal M}_{22} \right)^2 +
4 {\cal M}_{12}^2 } \right] \, ,
\ee
\be
\left(
\begin{array}{c}
V \\ V'
\end{array}
\right)
=
\left(
\begin{array}{cc}
c_{\alpha}
&
s_{\alpha}
\\
- s_{\alpha}
&
c_{\alpha}
\end{array}
\right)
\left(
\begin{array}{c}
V_1 \\ V_2
\end{array}
\right) \, ,
\ee
where $c_{\alpha} \equiv \cos \alpha$ and $s_{\alpha} \equiv 
\sin \alpha$, with
\be
\sin 2 \alpha =
{- 2 {\cal M}_{12} \over
\sqrt{ \left( {\cal M}_{11} - {\cal M}_{22} \right)^2 +
4 {\cal M}_{12}^2 } } \, ,
\;\;\;\;\;
\cos 2 \alpha =
{ {\cal M}_{22} - {\cal M}_{11} \over
\sqrt{ \left( {\cal M}_{11} - {\cal M}_{22} \right)^2 +
4 {\cal M}_{12}^2 } } \, .
\ee
In the limit of small mixing, $|\alpha| \ll 1$, and assuming
${\cal M}_{22} > {\cal M}_{11}$, but not necessarily
${\cal M}_{22} \gg {\cal M}_{11}$:
\be
\alpha \simeq
{{\cal M}_{12} \over
{\cal M}_{11} - {\cal M}_{22} }
\, ,
\ee
\be
V = V_1 + \alpha V_2 \, ,
\;\;\;\;\;
V' = V_2 - \alpha V_1 \, ,
\ee
\be
m_V^2 \simeq {\cal M}_{11} + {\cal M}_{12} \alpha \, ,
\;\;\;\;\;
m_{V'}^2 \simeq {\cal M}_{22} - {\cal M}_{12} \alpha \, .
\ee

\subsubsection{Vector Bosons}

In the charged vector boson sector, and in the $(W_L,W_R)$ basis:
\be
{\cal M}_{\pm}^2 =
{g^2 \over 4}
\left(
\begin{array}{cc}
v_L^2 + u^2 
&
- x u^2 \sin 2 \beta
\\
- x u^2 \sin 2 \beta
&
x^2 (v_R^2 + u^2)
\end{array}
\right)
\, .
\ee
In the limit of small mixing, as defined above, and in obvious 
notation:
\be
\alpha_{\pm} \simeq 
{x u^2 \sin 2 \beta \over x^2 (u^2 + v_R^2) - (u^2 + v_L^2)}
\, ,
\ee
\be
m_W^2 \simeq  {g^2 \over 4} \left[ (u^2 + v_L^2 )
- {x^2 u^4 \sin^2 2 \beta \over x^2 (u^2+v_R^2) - (u^2+v_L^2)}
\right]
\, ,
\ee
\be
m_{W'}^2 \simeq  {g^2 \over 4} \left[ x^2 (u^2 + v_R^2 )
+ {x^2 u^4 \sin^2 2 \beta \over x^2 (u^2+v_R^2) - (u^2+v_L^2)}
\right]
\, .
\ee

In the neutral  sector, and in the $(W^3_L,W^3_R,\tilde{B})$
basis
\be
{\cal M}_0^2 =
{g^2 \over 4}
\left(
\begin{array}{ccc}
v_L^2 + u^2
&
- x u^2
&
- y v_L^2
\\
- x u^2
&
x^2 (v_R^2 + u^2)
&
- x y  v_R^2
\\
- y v_L^2
&
- x y v_R^2
&
y^2 (v_L^2 + v_R^2)
\end{array}
\right)
\, .
\ee
It is convenient to move to the basis defined by
\be
\left(
\begin{array}{c}
A
\\
Z_L
\\
Z_R
\end{array}
\right)
=
U
\left(
\begin{array}{c}
W_L^3
\\
W_R^3
\\
B
\end{array}
\right)
\, ,
\ee
where
\be
U
=
\left(
\begin{array}{ccc}
{x y \over \sqrt{x^2+y^2+x^2y^2}}
&
{y \over \sqrt{x^2+y^2+x^2y^2}}
&
{x \over \sqrt{x^2+y^2+x^2y^2}}
\\
{\sqrt{x^2+ y^2} \over \sqrt{x^2+y^2+x^2y^2}}
&
- {x y^2 \over \sqrt{x^2+y^2} \sqrt{x^2+y^2+x^2y^2}}
&
- {x^2 y \over \sqrt{x^2+y^2} \sqrt{x^2+y^2+x^2y^2}}
\\
0
&
{x \over \sqrt{x^2+y^2}}
&
- {y \over \sqrt{x^2+y^2}}
\end{array}
\right)
\, .
\ee
In the $(A,Z_L,Z_R)$ basis, the mass matrix becomes block-diagonal:
\be
U {\cal M}_0^2 U^T =
\left(
\begin{array}{cc}
0 & \begin{array}{cc} 0 & 0 \end{array} \\
\begin{array}{c} 0 \\ 0 \end{array} & \tilde{{\cal M}}_0^2
\end{array}
\right) \, ,
\ee
and we can identify the photon with the massless combination $A$.
The non-vanishing block $\tilde{{\cal M}}_0^2$ is given by:
$$
\tilde{{\cal M}}_0^2 =
{g^2 \over 4 (x^2+y^2)}
$$
\be
\times
\left(
\begin{array}{cc}
(u^2+v_L^2)(x^2 +y^2+x^2y^2)
 &
(v_L^2 y^2 - u^2 x^2) \sqrt{x^2 + y^2 + x^2 y^2}
\\
(v_L^2 y^2 - u^2 x^2) \sqrt{x^2 + y^2 + x^2 y^2}
&
x^4 (u^2+v_R^2) + 2 x^2 y^2 v_R^2 + y^4 (v_L^2 + v_R^2)
\end{array}
\right) \, .
\ee
Working in the limit of small mixing, as defined above:
\be
\alpha_0 \simeq
{(v_L^2 y^2 - u^2 x^2) \sqrt{x^2 + y^2 + x^2 y^2}
\over (u^2+v_L^2)(x^2 +y^2+x^2y^2) -
x^4 (u^2+v_R^2) - 2 x^2 y^2 v_R^2 - y^4 (v_L^2 + v_R^2)}
\, ,
\ee
$$
m_Z^2 \simeq
{g^2 \over 4}{(x^2 +y^2+x^2y^2) \over (x^2+y^2)}
\left[ (u^2+v_L^2) \right.
$$
\be
+
\left.
{(v_L^2 y^2 - u^2 x^2)^2
\over (u^2+v_L^2)(x^2 +y^2+x^2y^2) -
x^4 (u^2+v_R^2) - 2 x^2 y^2 v_R^2 - y^4 (v_L^2 + v_R^2)}
\right] \, ,
\ee
$$
m_{Z'}^2 \simeq
{g^2 \over 4 (x^2+y^2)}
\left[x^4 (u^2+v_R^2)+2 x^2 y^2 v_R^2+y^4 (v_L^2 +v_R^2)  \right.
$$
\be
-
\left.
{(x^2+y^2+x^2y^2) (v_L^2 y^2 - u^2 x^2)^2
\over (u^2+v_L^2)(x^2 +y^2+x^2y^2) -
x^4 (u^2+v_R^2) - 2 x^2 y^2 v_R^2 - y^4 (v_L^2 + v_R^2)}
\right] \, .
\ee

\subsubsection{Fermions}

Fermion masses arise exactly as in the SM, via the Yukawa
interactions of eq.~(\ref{yuk}), involving fermion bilinears 
and the scalar 
doublet $\phi_L$ (we recall that in the FP model the scalar 
fields $\phi_R$ and $\phi_{LR}$ cannot have gauge-invariant 
couplings to fermion bilinears), thus they do not deserve any 
special discussion. The only point to notice is that, since 
fermion masses depend only on $v_L$, but gauge boson masses 
depend on all the four VEVs $(v_1,v_2,v_L,v_R$), the SM 
one-to-one correspondence between the numerical values 
of the fermion masses and the magnitude of the corresponding
Yukawa couplings is corrected by suitable mixing parameters.

\subsubsection{Scalars}

A complete description of the mass spectrum and of the interactions
in the scalar sector would require an explicit form of the scalar 
potential $V_0$ and its expansion around the minimum. Nevertheless, 
a parametrization for the mass spectrum and a discussion of some of 
its features can be outlined even in the absence of an explicit form
for $V_0$. Notice first that, out of the 16 spin--0 real degrees of 
freedom, 8 charged and 8 neutral, 6 (the would-be Goldstone bosons)
are absorbed as longitudinal components of the massive gauge bosons: 
these states can be unambiguously identified in terms of the 
components of the multiplets $\phi_L,~\phi_R,~\phi_{LR}$ and of 
the assumed pattern of VEVs. The remaining 10 degrees of freedom, 
4 charged and 6 neutral, correspond to physical spin--0 particles. 

In the charged Higgs sector, the physical mass eigenstates 
$H_{1,2}^\pm$ can be characterized by their two masses $m_{1,2}^\pm$
and by a single mixing angle $\beta_\pm$. Calling $G^\pm$ and $G'^\pm$ 
the charged would-be Goldstone bosons associated with $W$ and $W'$,
respectively, we can describe the relation among $\phi^{\pm} \equiv 
(\phi_R^\pm,\phi_L^\pm,\phi_1^\pm,\phi_2^\pm)^T$ and $H^{\pm} \equiv 
(G^\pm,G'^\pm,H_1^\pm,H_2^\pm)^T$ by a matrix equation:
\be
\label{rotcha}
\phi^\pm = A 
\cdot H^\pm \, ,
\ee
where the explicit form of the $4 \times 4$ orthogonal matrix A is
given in appendix~A.

If we assume no other sources of CP-violation besides the
Kobayashi-Maskawa phase, and in particular real parameters in 
the scalar potential
and real VEVs, the 6 physical states of the neutral Higgs sector can 
be divided into 2 CP--odd ($H_1^0,H_2^0$) and four CP-even states, and 
there is no mixing between the two sets. Collecting the physical CP-odd 
states and the neutral would-be Goldstone bosons ($G^0,G'^0$), associated
with the neutral gauge boson mass eigenstates $(Z,Z')$, in a vector 
$H^0 \equiv (G^0,G'^0,H_1^0,H_2^0)^T$, we can relate the mass
eigenstates $H^0$ with the interaction eigenstates $Im\phi^0 \equiv 
Im(\phi_R^0,\phi_L^0,\phi_1^0,\phi_2^0)^T$ in the following way:
\be
\label{rotneu}
\sqrt{2}\cdot Im\phi^0 = C 
\cdot H^0 \, ,
\ee
where the explicit form of the $4 \times 4$ orthogonal matrix $C$ is 
given in appendix~A. The neutral CP-odd sector is thus described by 
the two masses $m_{1,2}^0$ of $H_{1,2}^0$ and by a single mixing 
angle $\beta_0$.

Finally, the neutral CP--even states $Re\phi^0 \equiv Re(\phi_R^0,
\phi_L^0,\phi_1^0,\phi_2^0)^T$ can mix (6 mixing angles) to give 4 
mass eigenstates with masses $m_i^0,~(i=3,...6)$. We will not give 
here the general parametrization for this sector. As a zeroth-order 
approximation, we can identify the candidate SM-like Higgs, which 
is bound to survive as an approximate light mass eigenstate when
the scale of $SU(2)_R$ breaking and the remaining Higgs masses are
pushed much above the electroweak scale. To do so, we identify 
three $SU(2)_L$ doublets with identical SM quantum numbers
\be
\label{doublets}
\phi_L \equiv \left( \begin{array}{c} \phi_L^0 \\ \phi_L^-
\end{array} \right) \, ,
\;\;\;\;\;
\phi_L^1 \equiv \left( \begin{array}{c} \phi_1^0 \\ \phi_1^-
\end{array} \right)  \, ,
\;\;\;\;\;
\phi_L^2 \equiv i \sigma^2 \left( \begin{array}{c} \phi_2^+ \\
\phi_2^0
\end{array} \right) = \left( \begin{array}{c} \phi_2^{0\, *} \\
- \phi_2^- \end{array} \right) \, .
\ee
It is easy to identify the two-dimensional subspace of linear
combinations, $\chi^0 = \alpha {\, \rm Re \,} \Phi_L^0 + 
\beta {\, \rm Re \,} \Phi_1^0 + \gamma {\, \rm Re \,} \Phi_2^0$, 
with vanishing VEVs, $\langle \chi^0 \rangle = 0$. 
The orthogonal linear combination, appropriately normalized, 
will define the SM-like Higgs $h$:
\be
\label{smhiggs}
h = { \left[ v_L ( \sqrt{2}Re \, \phi_L^0 - v_L ) +
v_1 (\sqrt{2} Re \, \phi_1^0 - v_1 ) + v_2 ( \sqrt{2} Re \, \phi_2^0 - v_2 )
\right] \over \sqrt{v_L^2 + v_1^2 + v_2^2} } \, .
\ee
In the following sections, we shall often make the assumption that
$h$ is the only light mass eigenstate in the neutral Higgs 
boson sector.

\subsection{Interactions}

\subsubsection{Gauge interactions of fermions}

Charged-current gauge interactions of fermions are described by
\be
\label{chgau}
{\cal L}_{CC} = \left( \begin{array}{cc} J_L^{\mu +} & 0
\end{array} \right) \left( \begin{array}{c}  W_{\mu  \, L}^-
\\ W_{\mu \, R}^- \end{array} \right) + {\rm h.c.}
= \left( \begin{array}{cc} J_W^{\mu +} & J_{W'}^{\mu  +}
\end{array} \right) \left( \begin{array}{c}  W_{\mu}^-
\\ {W'_{\mu}}^- \end{array} \right) + {\rm h.c.} \, ,
\ee
where
\be
\label{ccurr}
J_W^{\mu +} = \cos \alpha_{\pm} J_L^{\mu +} \, ,
\;\;\;
J_{W'}^{\mu +} = - \sin \alpha_{\pm} J_L^{\mu +} \, ,
\ee
and, working with quark mass eigenstates and leaving implicit 
the generation indices, the charged current associated with 
$SU(2)_L$ fermion interactions is given by:
\be
\label{chcur}
J^{\mu +}_L  = {g \over \sqrt{2}} \left(
\overline{e_L} \gamma^{\mu} \nu_L +  \overline{d_L}
\gamma^{\mu} K^{\dagger} u_L \right) \, .
\ee
The Fermi coupling constant, as defined at the tree level from 
muon decay, is given by\footnote{Since the couplings of the 
charged Higgs bosons to fermions are proportional to the 
relevant fermion masses, given the constraints coming from 
heavy-flavour decays, to be discussed in section~3, we can 
safely neglect the charged-Higgs contributions to muon decay.}
\be
\dd{G_F \over \sqrt{2}} = \dd{g^2 \over 8}
\left( \dd{\cos^2 \alpha_{\pm} \over m_W^2} 
+ \dd{\sin^2 \alpha_{\pm} \over m_{W'}^2}
\right) \, .
\label{fermi}
\ee

Neutral current gauge interactions of fermions are described by
\be
{\cal L}_{NC} = \left( \begin{array}{ccc} J_L^{\mu 3} & 0
&  J_B^{\mu} \end{array} \right) \left( \begin{array}{c}
W_{\mu  \, L}^3 \\ W_{\mu \, R}^3 \\ B_{\mu} \end{array} \right)
\nn
\ee
\be
= \left( \begin{array}{ccc} J_{em}^{\mu} & J_{L}^{\mu  0} &
J_{R}^{\mu  0} \end{array} \right) \left( \begin{array}{c}
A_{\mu} \\ Z_{\mu}^L \\ Z_{\mu}^R \end{array} \right)
= \left( \begin{array}{ccc} J_{em}^{\mu} & J_{Z}^{\mu} &
J_{Z'}^{\mu} \end{array} \right) \left( \begin{array}{c}
A_{\mu} \\ Z_{\mu} \\ Z'_{\mu} \end{array} \right) \, ,
\ee
where, denoting with $f_i \equiv \{ e_L,e_R,\nu_L,u_L,u_R,d_L,d_R \}$
the chiral projections of the fermion fields and leaving implicit
the generation indices:
\be
J_L^{\mu 3} = g \sum_i T_{3L}^i \overline{f_i} \gamma^{\mu} f_i \,
,
\;\;\;\;\;
J_B^{\mu} = \tilde{g} \sum_i \tilde{Y}^i \overline{f_i} \gamma^{\mu}
f_i \, ,
\ee
\be
\left( \begin{array}{ccc} J_{em}^{\mu} & J_{L}^{\mu  0} &
J_{R}^{\mu  0} \end{array} \right) = \left( \begin{array}{ccc}
J_L^{\mu 3} & 0 &  J_B^{\mu} \end{array} \right) U^T \, ,
\ee
and
\be
J_Z = \cos \alpha_0 J_L^0 + \sin \alpha_0 J_R^0 \, ,
\;\;\;\;\;
J_{Z'} = - \sin \alpha_0 J_L^0 + \cos \alpha_0 J_R^0 \, .
\ee
It is convenient to write the explicit expression of the
electromagnetic current:
\be
J_{em}^{\mu} = e \sum_i Q^i \overline{f_i} \gamma^{\mu} f_i \, ,
\ee
where
\be
e 
= g {x y \over \sqrt{x^2+y^2+x^2y^2}} \, .
\ee
Notice that one recovers the SM tree-level relation $e = g \sin 
\theta_W$ by defining $\sin \theta_W$ as
\be
\sin \theta_W \equiv {xy\over\sqrt{x^2+y^2+x^2y^2}} \, .
\label{essew}
\ee 
Observe also the following simple relations:
\be
{1 \over e^2} = {1 \over g_R^2} + {1 \over g^2} + {1 \over
\tilde{g}^2}
\, , \;\;\;\;\;
A_{\mu} = {e \over g} W_{L \, \mu}^3 + {e \over g_R} 
W_{R \, \mu}^3 + {e \over \tilde{g}} B_{\mu} \, .
\ee
The remaining two neutral currents are given by:
\be
J_{L}^{\mu  0} = {g \over \cos \theta_W} \sum_i \left( T_{3L}^i -
Q^i \sin^2 \theta_W \right) \overline{f_i} \gamma^{\mu} f_i \, ,
\label{jzerol}
\ee
and
\be
J_{R}^{\mu  0} = - {g y^2 \over 
\sqrt{x^2 + y^2}} \sum_i \tilde{Y}^i  
\, \overline{f_i} \gamma^{\mu} f_i
\, . 
\ee
Notice that, due to the fermiophobic nature of $SU(2)_R$, 
the two charged currents $(J_W^{\mu +},J_{W'}^{\mu +})$
and the three neutral currents $(J_{em}^{\mu },J_{Z}^{\mu},
J_{Z'}^{\mu})$ are not linearly independent, in contrast 
with the other models of table~1. 

\subsubsection{Other interactions}

We comment now on other interaction terms that will be relevant
in the discussion of the model phenomenology.

Trilinear gauge boson vertices are completely determined by gauge
invariance and by the mixing angles in the gauge boson sector.
Their explicit expressions in the mass eigenstate basis are
collected in appendix~A.

The interaction terms involving the SM-like Higgs boson $h$ 
can be deduced from eqs.~(\ref{ls}) and (\ref{yuk}) by using 
eq.~(\ref{smhiggs}). In particular, the Yukawa interactions of 
$h$ have exactly the same form as for the SM Higgs, and the 
model shares with the SM the important property that there are 
no tree-level FCNC induced by the scalar sector:
\be
{\cal L}_Y^{h} = 
- {1 \over \sqrt{v_L^2 + v_1^2 + v_2^2 }} 
\left( 
\ov{u_L} M^U_{diag} u_R +
\ov{d_R} M^D_{diag} d_L +
\ov{e_R} M^E_{diag} e_L \right) h
+ {\; \rm h.c.}
\ee
It is also interesting to look at the interaction terms linear in 
$h$ and bilinear in the gauge boson mass eigenstates: their explicit 
expressions have been collected in appendix~A.

As for the Yukawa interactions of the physical charged Higgs bosons,
they may play a role in some decays of heavy flavours, such as
$b \rightarrow c \tau^- \ov{\nu}_{\tau}$ or $t \rightarrow b
H_i^+$, as well as in the generation of FCNC and of non-standard
contributions to $\Gamma(Z^0 \rightarrow b \ov{b})$ at the one-loop 
level. In view of the following discussion, it may be useful to 
rewrite these interactions in terms of the physical charged Higgs 
mass eigenstates, defined in eq.~(\ref{rotcha}):
\be
\label{chyuk}
{\cal L}_Y^{ch} = 
{\sqrt{2} \over v_L} \left( 
- \ov{d_L} K^{\dagger} M^U_{diag} u_R 
+ \ov{d_R} M^D_{diag} K^{\dagger}  u_L
+ \ov{e_R} M^E_{diag} \nu_L \right) 
\sum_{i=3,4} A_{2i} H_i^- + {\; \rm h.c.}
\ee
In the following section, we shall often consider the limiting
case of vanishing mixing angle in the charged gauge boson sector
and of degenerate physical charged Higgs bosons:
\be
\label{limcase}
\alpha_{\pm} = 0 \, ,
\;\;\;\;\;\;\;\;\;
m_1^{\pm} = m_2^{\pm} = m_H \, .
\ee
In such a case, when dealing with processes controlled by
the Yukawa interactions of the physical charged Higgs bosons, 
we can forget about the mixing angle $\beta_{\pm}$ and work 
as if there were a single charged Higgs boson, $H^{\pm}$, with
\be
\label{chyukbis}
{\cal L}_Y^{ch} = -
{g \over \sqrt{2} m_W} 
{\tan \theta_W \over \sqrt{x^2 - \tan^2 \theta_W}}
\left( 
- \ov{d_L} K^{\dagger} M^U_{diag} u_R 
+ \ov{d_R} M^D_{diag} K^{\dagger}  u_L
+ \ov{e_R} M^E_{diag} \nu_L \right) H^- 
+ {\; \rm h.c.}
\ee
\nsect{Phenomenology of the fermiophobic model}

\subsection{Approximate parametrization}

Considering for the moment only gauge interactions, the model 
has 7 independent parameters, three gauge couplings $(g, g_R,
\tilde{g})$ and four VEVs $(v_1,v_2,v_L,v_R)$. However, it is 
convenient to move to suitable combinations of these parameters 
with a more direct physical interpretation. To replace the gauge 
couplings, we choose the electric charge $e \equiv \sqrt{4 \pi 
\alpha}$, the electroweak mixing angle $\theta_W$ and the ratio 
$x\equiv g_R/g$ between the two non-abelian couplings. The 
exact translation table is:
\be
\label{transl}
g = {e \over s_W} \, ,
\;\;\;\;
g_R = { e x \over s_W} \, ,
\;\;\;\;
\tilde{g} = { e x \over \sqrt{x^2 c_W^2 - s_W^2} }
\, ,
\ee
where $s_W \equiv \sin \theta_W$ and $c_W \equiv \cos \theta_W$. 
Notice that the physical
requirement $\tilde{g}^2 > 0$ corresponds to the constraint
$x > \tan \theta_W \simeq 0.55$. To replace the four VEVs, we 
choose two gauge boson masses, for example $m_Z$ and $m_{W'}$, 
and the two mixing angles, $\alpha_{\pm}$ and $\alpha_0$, assumed
to be small\footnote{The allowed range for the mixing angles 
$\alpha_{\pm}$ and $\alpha_0$ is not completely arbitrary:
once ($x,m_{W'}$) are given, and $(\alpha,m_Z,\sin \theta_W)$
extracted from experiment, $|\alpha_{\pm}|$ and $|\alpha_0|$
are bounded from above, as can be seen by inspecting the
vector boson mass matrices.}. We can then derive simple
approximate expressions for the other relevant quantities in
the gauge sector. At the lowest non-trivial order in the 
mixing angles, i.e. neglecting ${\cal O} ( \alpha_{\pm,0}^2 )$
terms, we find\footnote{We should warn the reader that, for 
very large values of $x$ or $m_{W'}$, subleading terms in the 
$(\alpha_0,\alpha_\pm)$ expansion may become non-negligible. For 
our numerical results we will always use the complete formulae.}:
\be
m_W^2 \simeq c_W^2 m_Z^2 \, ,
\;\;\;\;\;\;
m_{Z'}^2 \simeq { x^2 c_W^2 m_{W'}^2 - s_W^4 m_Z^2
\over x^2 c_W^2 - s_W^2 } \, ,
\ee
\be
u^2 \equiv v_1^2 + v_2^2 \simeq {s_W^4 \over \pi \alpha
x^2} m_Z^2 \, ,
\;\;\;\;\;\;
\sin 2 \beta \simeq x {m_{W'}^2 - c_W^2 m_Z^2 \over
s_W^2 m_Z^2} \alpha_{\pm} \, ,
\ee
\be
v_L^2 \simeq {s_W^2 ( x^2 c_W^2 - s_W^2) \over
\pi \alpha x^2} m_Z^2 \, ,
\;\;\;\;\;\;
v_R^2 \simeq  {s_W^2 \over \pi \alpha x^2}
(m_{W'}^2 - s_W^2 m_Z^2) \, .
\ee

When dealing with precision tests of the model, we must be 
more precise in our definitions of the input parameters. In 
particular, it is convenient to express the electroweak 
mixing angle $\theta_W$ in terms of ($G_F,\alpha,m_Z$). 
From eqs.~(\ref{transl}) and (\ref{fermi}), we get
\be
m_W^2={\mu^2\over s_W^2} \left[1+\sin^2\alpha_\pm
\left({m_W^2-m_{W'}^2\over m_{W'}^2}\right)\right]\,  ,
\label{mw2}
\ee
where
\be
\mu^2 \equiv {\pi\alpha\over\sqrt{2} G_F}\, .
\ee
Notice that, at the lowest non-trivial order in the mixing,
the relation among $G_F$, $\alpha$, $m_W$ and $s_W$ remains 
the same as in the SM. To eliminate $m_W$ in favour of $m_Z$ 
in eq.~(\ref{mw2}), we proceed as in the SM, defining
\be
\rho \equiv {m_W^2 \over m_Z^2 c_W^2} \, .
\label{rho}
\ee
In the limit of small mixing, we find:
\be
\rho \equiv 1 +  \Delta\rho = 
1 + \Delta \rho_W + \Delta \rho_Z \, ,
\ee
where
\be
\label{drhowz}
\Delta \rho_W \simeq - \alpha_{\pm}^2
{m_{W'}^2 - m_W^2 \over m_W^2} \, ,
\;\;\;\;\;
\Delta \rho_Z \simeq   \alpha_0^2
{m_{Z'}^2 - m_Z^2 \over m_Z^2} \, .
\ee
By combining eqs.~(\ref{mw2}) and (\ref{rho}), we obtain:
\be
s_W^2\simeq\bar{s}^2-{\bar{c}^2\bar{s}^2\over(\bar{c}^2
-\bar{s}^2)}\Delta\rho_{eff}\, ,
\label{sw2}
\ee
where
\be
\bar{s}^2={1\over 2}-\sqrt{{1\over 4}-{\mu^2\over m_Z^2}}
\ee
corresponds to the well-known tree-level SM relation, and
\be
\label{drho}
\Delta\rho_{eff} \simeq \Delta\rho-\alpha_\pm^2 
\left({m_W^2-m_{W'}^2\over m_{W'}^2}\right)
\simeq \alpha_0^2 {m_{Z'}^2 - m_Z^2 \over m_Z^2}-
\alpha_{\pm}^2 {(m_{W'}^2 - m_W^2)^2 \over m_W^2 
m_{W'}^2}
\ee
parametrizes the deviation from it, still at the classical level. 
Eq.~(\ref{sw2}) allows to express $\theta_W$ in terms of the 
input parameters ($G_F,\alpha,m_Z$), plus corrections vanishing 
in the limit of zero mixing angles. In summary, we can use as 
independent parameters ($G_F,\alpha,m_Z$), the same input 
quantities of the SM precision tests, plus $\alpha_\pm$, 
$\alpha_0$, $m_{W'}$ and $x$. 

There are two combinations of these parameters which are 
particularly relevant to our analysis. The first one is the 
ratio $g/c_W$ appearing in the expression of the neutral 
current $J^0_L$, eq.~(\ref{jzerol}). In the limit of small 
mixing angle:
\be
{g^2\over c_W^2}=4\sqrt{2} G_F m_Z^2 (1+\Delta\rho_{eff})\, .
\ee
This relation shows that the strength of the neutral current 
gets corrected by the same parameter, $\Delta\rho_{eff}$,
that modifies the weak mixing angle $\theta_W$. Of course, 
quantum corrections also modify the classical SM relations 
and, in particular, may contribute to the parameter $\Delta
\rho_{eff}$ of eq.~(\ref{drho}): we will discuss this issue 
later on. The second quantity of interest is the ratio $m_W
/m_Z$, which satisfies the relation:
\be
{m_W^2\over m_Z^2} \left(1-{m_W^2\over m_Z^2}\right)={\mu^2\over m_Z^2 
(1-\Delta r)} \, ,
\ee
with
\be
\Delta r \simeq -{\bar{c}^2\over\bar{s}^2} \Delta\rho
+\alpha_\pm^2\left({m_W^2-m_{W'}^2\over m_{W'}^2}\right) \, .
\ee

In the remaining sections we will allow the parameter $x$ to vary 
in a wide range, starting from its lower bound, $\tan\theta_W$, 
up to values as large as 20. We may wonder about the properties 
of the theory in the large $x$ regime (similar considerations
have been recently made, in a similar context, in ref.~\cite{casal}). 
In particular, we would like to maintain control over the predictions 
that are relevant to our analysis, even in the presence of the strong 
interactions associated with $g_R$. It turns out that, when $x$ is 
large, the states of the model split into two sectors. The first 
sector includes the new vector bosons $W'$ and $Z'$ and the scalar 
mass eigenstates having projections along the multiplets $\phi_R$ and 
$\phi_{LR}$. This sector experiences the strong interaction related 
to the large $g_R$ coupling. The second sector comprises the ordinary 
vector bosons, the fermions and the rest of the scalar sector. The 
interactions among these states do not grow with $x$, at least in the 
case of vanishing mixing angles $\alpha_0$ and $\alpha_\pm$, which 
represents, as we shall see, a good approximation to the realistic 
case. Finally, when considering processes involving only ordinary
particles, which belong to the second sector, the corrections 
induced by the states of the first sector are bounded in the large 
$x$ limit, for vanishing $\alpha_0$ and $\alpha_\pm$.
This structure guarantees that, as long as we work at energies
below the threshold of particle production in the first sector,
the strong interaction cannot propagate to the states of the
second sector. We will sometimes consider 
the possibility of producing real $W'$ and $Z'$. In this case a 
very large value of $x$ might lead to violation of perturbative 
unitarity. We restrict our numerical analysis to $x \simlt 20$, 
corresponding to $g_R \simlt 13$.

\subsection{Tree-level fit to electroweak data}

A first important constraint on the parameter space of the 
model comes from the comparison with the electroweak data 
collected at the $Z$ peak, the ratio $m_W/m_Z$ and the 
low-energy data from neutrino-hadron scattering and atomic 
parity violation experiments. A recent compilation of these 
data \cite{data} is shown in table~3.
\begin{table}[htb]
\begin{center}
\begin{tabular}{|c|c|}   
\hline                        
& \\
Quantity &  Exp. values \\
& \\                                                  
\hline\hline
&  \\
${\Gamma_Z} ({\rm MeV})$   
& $2494.7 \pm 2.6$  
\\                  
$R_l = \Gamma_h/\Gamma_l$  
& $20.783 \pm 0.029$ 
\\        
$\sigma_h (nb)  $ 
& $41.489 \pm 0.055$ 
\\
$R_b = \Gamma_b/\Gamma_h$                
& $0.2177 \pm 0.0011$ 
\\                                                          
$R_c = \Gamma_c/\Gamma_h$ 
& $0.1722 \pm 0.0053$ 
\\
$m_W/m_Z$ 
& $0.8814 \pm 0.0008$ 
\\                     
${\cal A}_l$  
& $-0.1512 \pm 0.0023$ 
\\                 
${\cal A}_b$ 
& $-0.897 \pm 0.047$ 
\\
${\cal A}_c$ 
& $-0.623 \pm 0.085$ 
\\
$A_{FB}^{b}$ 
& $0.0985 \pm 0.0022$ 
\\
$A_{FB}^{c}$ 
& $0.0735 \pm 0.0048$ 
\\
$Q_{W}(Cs)$ 
& $-72.08 \pm 0.93$ 
\\
$g^2$ & $0.3017 \pm 0.0033$ 
\\
$g_R^2$ 
&$0.0326\pm 0.0033$ 
\\
& \\
\hline                        
\end{tabular}
\end{center}
\caption{{\it Experimental values for the electroweak observables
used in our fit.}}
\end{table}
In general, the deviation from the SM prediction of the generic 
observable of table~3 depends on the parameters $\alpha_0$, 
$\alpha_\pm$, $x$ and $m_{W'}$. The main dependence comes through 
the combination $\Delta\rho_{eff}$, which modifies both the 
electroweak mixing angle $\theta_W$ and the strength of the neutral 
current, and through the combination $\alpha_0 y^2/\sqrt{x^2+y^2}$, 
which controls the amount of contamination of the ordinary neutral 
current $J^0_L$ by the new current $J^0_R$. Exceptions to this rule 
are the low-energy observables $g_{L,R}^2$, associated with
neutrino-hadron scattering (and not to be confused with the $SU(2)$
coupling constants) and $Q_W$, associated with atomic parity violation, 
which are also
affected by direct $Z'$ exchange, and the ratio $m_W/m_Z$, which
is corrected by $\Delta r$. The explicit form of these corrections
can be easily obtained, following for instance the procedure outlined 
in refs.~\cite{zp}, and will not be reported here. To test the model 
against the electroweak data, we have performed a fit to the 14 
observables of table~3. The theoretical predictions of the model have 
been obtained by adding to the SM predictions, radiative corrections
included, the appropriate deviations, as computed at the classical
level in the FP model. A $\chi^2$ minimization procedure determines 
the best values and the errors for the parameters of the fit, to be 
chosen among $\alpha_0$, $\alpha_\pm$, $x$ and $m_{W'}$. Besides the 
input values for ($G_F,\alpha,m_Z$), the SM one-loop predictions also 
depend on the top mass $m_t$, the Higgs mass $m_h$ and the strong 
coupling constant $\alpha_s$, which will be kept fixed. The Higgs 
boson $h$ is identified here with what we defined as SM-like 
Higgs in eq.~(\ref{smhiggs}). Additional dependences of the radiative 
corrections upon the other scalar particles and the additional gauge
bosons will be addressed separately in the following section. 

To keep the number of fit parameters reasonably small, we fix the $x$ 
parameter by choosing the following set of representative values: 0.6,
1, 2.5, 5, 15. Then we observe that the fit is quite insensitive to the 
$W'$ mass: $m_{W'}$ is determined with a huge error. In view of this, 
we prefer to keep also $m_{W'}$ fixed in the minimization procedure, 
and we assign to it some representative values in the range 100--1000 
GeV. The final results are displayed in tables 4 and 5, where we report 
the best values and the $1 \sigma$ errors for $\alpha_0$ ($\alpha_\pm$) 
in units of $10^{-3}$.
\begin{table}[htb]
\begin{center}
\begin{tabular}{|c|c|c|c|c|c|}
\hline
$
\begin{array}{c} 
m_{W'} 
\\ 
{\rm (GeV)}
\end{array}
$
& $x=0.6$ & $x=1$ & $x=2.5$ & $x=5$ & $x=15$
\\
\hline
100 & - & - & - & $-11.3 \pm 5.2$ & $-34.7 \pm 13.0$
\\
    & - & - & - &   ($0 \pm 29$)  &   ($0 \pm 37$) 
\\
\hline
200 & - & - & $-5.9 \pm 2.4$ & $-10.6 \pm 3.3 $ & $-23.7 \pm 15.7$
\\
    & - & - &   ($0 \pm 8$)  &  ($0 \pm 14$)    &($-18.0 \pm 18.9$) 
\\
\hline
500 & $-0.6 \pm 0.2$ & $-2.0 \pm 0.8$ & $-4.4 \pm 1.2$ & 
$-8.2 \pm 5.8$   & $-23.6 \pm 15.1$
\\
    &  ($0 \pm 3$)   & ($0 \pm 3$)    & ($0 \pm 10$)   & 
($-5.9 \pm 6.9$) &($-20.5 \pm 13.8$) 
\\
\hline
1000 & $-0.6 \pm 0.2$ & $-1.7 \pm 0.5$ & $-4.0 \pm 2.9$ & 
$-8.2 \pm 5.7$  & $-23.6 \pm 13.9$
\\
     & ($0 \pm 2$)    & ($0 \pm 3$)    &($-3.0 \pm 3.4$)& 
($-6.9 \pm 5.4$)    & ($-20.7 \pm 12.3$) 
\\
\hline
\end{tabular}
\end{center}
\caption{{\it Best values and 1$\sigma$ errors for $\alpha_0$ 
($\alpha_\pm$) in units of $10^{-3}$, for the indicated values 
of $x$ and $m_{W'}$, and: $m_t=175~{\rm GeV}$, $m_h=300~{\rm GeV}$,
$\alpha_s(m_Z)=0.118$. Here 0 stands for a value smaller than 
$10^{-6}$. Where no value is indicated, the minimum $\chi^2$ is 
larger than 25.}}
\end{table}
From table~4 we see that $\alpha_0$ scales approximately as $x$, which
confirms the fact that the deviations for the LEP observables, beyond
$\Delta\rho_{eff}$, depend only on the combination $\alpha_0 y^2/
\sqrt{x^2+y^2}$. We also notice that, most of the times, the
value of $\alpha_\pm$, although affected by a large error, 
is very close to zero. This can be understood in terms of the 
contribution to $\Delta\rho_{eff}$ proportional to $\alpha_\pm^2$.
This contribution, detailed in eq. (\ref{drho}), is always negative.
On the other hand, for the chosen values of $m_t$ and $m_h$,
the data require a positive $\Delta\rho_{eff}$ and force the 
$\alpha_\pm^2$ contribution to vanish. 

For small values of $x$, ($x<5$), $\alpha_0$ is also small, ${\cal 
O} (10^{-3})$, and its contribution to $\Delta\rho_{eff}$ remains
within the allowed experimental range even for very large $m_{Z'}$
values. On the contrary, for large values of $x$, ($x>5$), the best 
value of $\alpha_0$ is close to $10^{-2}$. In this case, when large
$m_{W'}$ or $m_{Z'}$ are considered, the positive $\alpha_0^2$ contribution 
to $\Delta\rho_{eff}$ is too large and a compensating negative
term is required. This explains why, for large $x$ and $m_{W'}$,
the preferred values for $\alpha_\pm$ are non-vanishing and approximately
equal in size to $\alpha_0$. 
\begin{table}[htb]
\begin{center}
\begin{tabular}{|c|c|c|c|c|c|}
\hline
$
\begin{array}{c} 
m_{W'} 
\\ 
{\rm (GeV)}
\end{array}
$
& $x=0.6$ & $x=1$ & $x=2.5$ & $x=5$ & $x=15$
\\
\hline
100 & - & - & - & $-0.5 \pm 5.7$ & $-1.2 \pm 17.0$
\\
    & - & - & - & ($-47.2 \pm 21.3$) & ($-47.0 \pm 21.4$)
\\
\hline
200 & - & - & $-1.0 \pm 2.9$    & $-1.9 \pm 5.9$   & $-5.6 \pm 17.8$
\\
    & - & - & ($-12.3 \pm 4.9$) & ($-12.3 \pm 5.2$) & ($-13.2 \pm 9.3$)
\\
\hline
500 & $-0.2 \pm 0.3$ & $-0.4 \pm 1.0$ & $-1.0 \pm 2.9$ & 
$-2.0 \pm 5.9$ & $-6.4 \pm 17.4$
\\
    & ($-4.4 \pm 1.6$) & ($-4.3 \pm 1.7$) & ($-4.4 \pm 1.9$) & 
($-4.6 \pm 3.0$) & ($-7.1 \pm 12.7$) 
\\
\hline
1000 & $-0.1 \pm 0.3$ & $-0.4\pm 1.0$ & $-1.0 \pm 2.9$ & 
$-2.0 \pm 5.8$ & $-6.5 \pm 15.7$
\\
     & ($-2.1 \pm 0.8$) & ($-2.1 \pm 0.9$) & ($-2.3 \pm 1.5$) & 
($-2.8 \pm 3.7$) & ($-6.1 \pm 13.1$) 
\\
\hline
\end{tabular}
\end{center}
\caption{{ \it Best values and 1$\sigma$ errors for $\alpha_0$ 
($\alpha_\pm$) in units of $10^{-3}$, for the indicated values 
of $x$ and $m_{W'}$, and: $m_t=175~{\rm GeV}$, $m_h=100~{\rm GeV}$,
$\alpha_s(m_Z)=0.118$. The $\chi^2$ minimum is between 17 and 25.
Where no value is indicated, the minimum $\chi^2$ is larger than 
25.}}
\end{table}

In table~5 we present the results for $m_t=175~{\rm GeV}$, 
$m_h=100~{\rm GeV}$ and $\alpha_s(m_Z)=0.118$. Notice that, 
for $W'$ masses in the chosen range, the fit is insensitive to 
$m_{W'}$. Only when $x$ is small, the $\chi^2$ minimum indicates 
that large values of $m_{W'}$ are preferred due to the potentially 
large contributions to the low-energy observables via direct $Z'$ 
exchange (this also happens in the case discussed before). Moreover, 
$\alpha_0$ is, to a large extent, independent of $m_{W'}$, and 
scales approximately with $x$. Finally, for $x<5$ the best value of 
$\alpha_\pm$ is insensitive to $x$ and is smaller for larger $m_{W'}$ 
values. Indeed, for the chosen values of $m_t$ and $m_h$, the SM 
contribution to $\Delta\rho$ tends to exceed the experimentally 
allowed one. This excess is in turn compensated by the negative 
contribution coming from $\Delta\rho_W$ of eq.~(\ref{drhowz}), with 
a suitable combination of $\alpha_\pm$ and $m_{W'}$. This also 
explains why the central value of $\alpha_\pm$ is non vanishing,
contrary to the case of table~4. When $x$ gets large ($x>5$), the 
situation is similar to that discussed for $m_h=300 {\; \rm GeV}$ 
and analogous considerations apply.

As explained above, in performing the fit we have only considered
the SM one-loop corrections, neglecting the radiative corrections
which may be originated by the additional sectors of the FP model 
coupled to the SM. The validity of such an approximation will be 
discussed in the next subsection. For the moment we can observe 
that, in the absence of new one-loop corrections quadratically 
dependent on combinations of particle masses, the numerical 
difference between the results of tables 4 and 5 may be regarded 
as indicative of the theoretical uncertainty underlying the present 
discussion.

In summary, comparison with electroweak precision data
allows for mixing angles ($\alpha_0,\alpha_\pm$) in the range 
$10^{-3}-10^{-2}$, depending on the value of $x$: larger mixing 
for larger $x$. The allowed mass range for $W'$ and $Z'$ is broad, 
and even relatively light new vector bosons can be acceptable, 
except for small values of $x$.

\subsection{One-loop corrections from the scalar sector}

In performing the fit described in the previous section, we have 
tacitly assumed that the most important quantum corrections to the
electroweak observables in the
FP model are the SM ones. In other words, we have neglected all 
the loop corrections due to the additional particles of the FP 
model. In view of the precision reached by the present
electroweak data, we would like to comment here about the 
possible validity of such an approximation, considering the
one-loop contributions to $\Delta \rho$ and to $R_b \equiv \Gamma_b
/ \Gamma_h$ due to the scalar sector of the FP model. 

\subsubsection{$\Delta \rho$}

In general, we expect that the most dangerous non--SM radiative
correction may be the one-loop contribution to $\Delta\rho$ 
due to the scalar sector of the FP model. Indeed, on dimensional 
grounds this contribution can depend quadratically upon the masses 
of the scalar particles, and even for masses within few hundred GeV, 
it can easily reach the by now intolerable percent level. Therefore,
it is important to look for those configurations of the scalar sector 
that could appropriately deplete such correction. It is not difficult
to figure out that such configurations indeed exist, in some 
particular limit of the model (for a generalization to larger groups,
see \cite{quim}). 

Consider for instance the large $v_R$ limit, at fixed values of the 
remaining parameters. In this limit, the $SU(2)_R$ symmetry is broken 
at a scale much higher than the $SU(2)_L\otimes U(1)_Y$ breaking scale, 
${G_F}^{-1/2}$, and it makes sense to consider an effective theory 
valid at energies close to ${G_F}^{-1/2}$. The effective theory
possesses an $SU(2)_L\otimes U(1)_Y$ gauge invariance and,
if the physical scalar contained in the $\phi_R$ multiplet
is an approximate mass eigenstate, with a mass of order $v_R$,
the surviving light scalar sector includes only the multiplets
$\phi_L$ and $\phi_{LR}$. From the point of view of the low-energy
gauge symmetry, these are just three $SU(2)_L$ doublets, with
the same hypercharge, which we have already denoted by ($\phi_L, 
\phi_L^1,\phi_L^2$) in eq.~(\ref{doublets}). Their linear combination 
with non-vanishing VEV, eq.~(\ref{smhiggs}), corresponds to the 
SM-like Higgs $h$, whilst the two orthogonal combinations are just 
additional matter multiplets with no r\^ole in the symmetry breaking 
mechanism. In particular, if the physical components of these 
multiplets are approximate mass eigenstates, then it is very simple 
to compute the corresponding one-loop contribution to the $\Delta\rho$ 
parameter. We obtain the usual SM Higgs contribution, which is at most 
logarithmic in the Higgs mass, plus the contribution of two extra scalar 
doublets, which reads:
\bea
\Delta \rho&=& {g^{2} \over 64 \pi^2 m_{W}^{2}} \left\{ 
m_{1}^{\pm \, 2} - g (m_{5},m_{1}^{0}) +  \left[ g (m_{1}^{\pm},m_{5}) 
+ g (m_{1}^{\pm},m_{1}^{0}) \right]\right.  \nn \\
& & \phantom{{g^{2} \over 64 \pi m_W}} + \left. m_{2}^{\pm \, 2}
- g (m_{6},m_{2}^{0}) + \left[
g (m_{2}^{\pm},m_{6}) + g (m_{2}^{\pm},m_{2}^{0}) \right] \right\}
\, ,
\eea
where
\be
g(m_{i},m_{j})={m_{i}^2 m_{j}^{2} \over m_{j}^2-m_{i}^2} \log
{m_{i}^{2} \over m_{j}^{2}} \, ,
\ee
and we have denoted by $m_5$ and $m_6$ the masses of the two neutral 
CP-even bosons that sit in the same doublets with $(H_1^\pm,~H_1^0)$
and $(H_2^\pm,~H_2^0)$, respectively. From the previous expression 
one recovers immediately the well-known result that, if there are 
no mass splittings ($m_{1}^{\pm}=m_{1}^{0}=m_{5}$ and $m_{2}^{\pm} =
m_{2}^{0}=m_{6}$) between the scalars inside each doublet, the 
quadratic contribution to $\Delta\rho$ vanishes for arbitrary 
values of the common scalar masses. We also notice that, in the 
large $v_R$ limit, the masses of the observed gauge bosons are 
given by:
\be \label{mas1}
m_{W}^{2}={g^2  \over 4} (v_{L}^{2} + u^{2}) \, , 
\;\;\;\;\;
m_{Z}^{2}={g^2 \over  4} (v_{L}^2 + u^{2}) 
{x^{2} + y^{2} + x^{2} y^{2} \over x^{2}+y^{2}} \, ,
\ee
and the tree level $\rho$ parameter is exactly equal to one.

Indeed, the above cancellation of
$\Delta\rho$, both at tree-level and in the one-loop approximation,
may be related to the same custodial symmetry \cite{custodial} that 
protects the $\rho$ parameter in the SM. In the considered limit, 
the multiplet $\phi_R$, containing the would-be Goldstone bosons 
absorbed by $W'$, $Z'$ and a neutral physical scalar, is a singlet 
of the custodial $SU(2)$ and does not affect the $\rho$ parameter. 
Of the remaining doublets, those with non vanishing VEVs are 
doublets of the custodial $SU(2)$ and should be degenerate to 
preserve $\rho=1$; the one acquiring a VEV splits in a triplet 
plus a singlet, as in the SM.

There are other configurations of the scalar sector that are reminiscent
of a custodial symmetry. We observe that, when $y=0$ and $v_1=v_2=v$,
the mass matrices in the neutral and in the charged sectors coincide.
In particular, in this simplified  situation, the photon corresponds
to the $B$ gauge boson and the massive gauge bosons are admixtures
of the $W_L^i$, $W_R^i$ states. Moreover, the mixing angles $\alpha_0$ 
and $\alpha_\pm$ are the same, which allows to discuss in simple terms 
the interactions of the physical $W$ and $Z$ with the scalar particles. 
To this purpose, it is instructive to express the covariant derivatives
acting on the scalar fields as functions of the mass eigenstates $W$ 
and $Z$:
\bea
D_\mu \phi_L&=&(\partial_\mu
-i g \cos\alpha {W^a}_\mu\dd\frac{\tau^a}{2}
)\phi_L +...\, , \nn\\
D_\mu \phi_R&=&(\partial_\mu
-i g x \sin\alpha {W^a}_\mu\dd\frac{\tau^a}{2}
)\phi_R+... \, , \nn\\
D_\mu \phi_{LR}&=&\partial_\mu\phi_{LR}
-i g \cos\alpha {W^a}_\mu\dd\frac{\tau^a}{2}\phi_{LR}
+i g x \sin\alpha {W^a}_\mu\phi_{LR}\dd\frac{\tau^a}{2} +...\, .
\eea
where we have denoted with $\alpha$ the common value $\alpha_0=\alpha_\pm$,
with $W^a_\mu$ the mass eigenstates ($W^\pm_\mu,Z_\mu$), and the dots stand
for terms containing the $W'$ and $Z'$ fields. Notice that, if $\cos\alpha
=x\sin\alpha$, then, as far as the scalar sector is concerned, 
it is possible to define an $SU(2)_W$ transformation
under which $\phi_L$ and $\phi_R$ transform as complex doublets, 
whereas $\phi_{LR}$ decomposes in a complex triplet $\phi_3$ plus a
complex singlet $\phi_1$.

It is also useful to think of $\phi_L$ and $\phi_R$ as doublets of 
an additional global $SU(2)_X$, acting on the right of the multiplets, 
when written as $2\times 2$ matrices:
\be
\phi_{L,R}\leftrightarrow
\left( \begin{array}{cc} \phi^0_{L,R} & -\phi^+_{L,R} \\ 
\phi^-_{L,R} & (\phi^0_{L,R})^* \end{array} \right)
\ee
The multiplets $\phi_3$ and $\phi_1$ are singlets under $SU(2)_X$.
The assumed pattern of VEV's breaks $SU(2)_W\otimes SU(2)_X$ down
to the diagonal subgroup $SU(2)_C$, which defines a custodial symmetry.

The multiplets with a non-vanishing VEV are now $\phi_L$, $\phi_R$
and, due to the $v_1=v_2=v$ condition, $\phi_1$. The would-be Goldstone
bosons eaten up by the massive $W$ and $Z$ are contained in the 
$\phi_L$, $\phi_R$ doublets, whereas those absorbed by the $W'$ and 
$Z'$ particles are generically shared by all the multiplets.
If we further require $v \gg v_L,v_R$, then the Goldstone modes related
to $W'$ and $Z'$ are only contained in $\phi_3\oplus \phi_1$.
The physical scalars in the spectrum, classified according to $SU(2)_C$,
are now: a complex doublet, linear combination of $\phi_L$ and 
$\phi_R$ with vanishing VEV, a neutral scalar belonging to the 
combination of $\phi_L$ and $\phi_R$ with non-vanishing VEV and 
singlet under $SU(2)_C$, a real triplet coming from $\phi_3$ after 
subtracting the Goldstones and, finally, a complex singlet $\phi_1$. 
This physical spectrum corresponds, in our conventions of appendix~A, 
to $\beta_\pm=\beta_0=0$. The overall one-loop contribution to
$\Delta\rho$ coming from the scalar sector will now be that of a 
complex doublet plus a real triplet of matter particles. An 
explicit computation gives:
\bea
\Delta\rho & = & {x^2\over 16 \pi^2 v_L^2 (1+x^2)} \left[ 
{m_1^\pm}^2 - g (m_{5},m_{1}^{0}) +  g (m_{1}^{\pm},m_{5}) 
\right. \nn \\ & + & \left. g (m_{1}^{\pm},m_{1}^{0}) 
+ 2({m_2^\pm}^2+{m_6}^2) + 4 g(m_2^\pm,m_6)\right] \, ,
\eea
where $m_5$ and $m_6$ are the masses of the neutral CP-even
states in the linear combination of $\phi_L$ and $\phi_R$
with vanishing VEV and in $\phi_3$, respectively. This 
contribution vanishes for degenerate multiplets, that is 
for $m_1^\pm=m_1^0=m_5$ and $m_2^\pm=m_6$.
Moreover, we have explicitly verified that, when $y\ne 0$, the
previous result may receive only logarithmic corrections, as is the case 
for the SM. In this case the mixing angles in the charged and neutral
sector are no longer the same and the condition $\cos\alpha=x \sin\alpha$
should be replaced by $v_R=v_L/x$. To summarize, when $v_1=v_2=v \gg
v_R=v_L/x$, one-loop corrections to the $\rho$ parameter 
quadratic in the scalar masses can be avoided, if the physical scalars
fit in appropriate degenerate multiplets of a custodial $SU(2)_C$.
In this configuration (before taking the large $v$ limit), 
the exact tree-level masses for the gauge bosons become particularly simple
\bea 
\label{mas2}
&&m_{W}^{2}={g^2 v_{L}^{2} \over 4} \, , \quad \quad
m_{Z}^{2}={g^2 v_{L}^{2} \over  4} {(x^{2}+y^{2}+ x^{2} y^{2}) 
\over x^{2}} \, ,  \nn \\
&&m_{W^{\prime}}^{2}=m_{Z^{\prime}}^{2}={g^{2} \over 4} (v_{L}^{2}+
2 v^{2} (1 + x^{2}) ) \; ,
\eea
and the mixing angles in the gauge boson sector are simply given by:
\be
\label{angles}
\tan \alpha_{\pm} = {1 \over x} \, ,
\;\;\;\;\;
\tan \alpha_0 = {1 \over \sqrt{x^2 c_W^2 - s_W^2}} \, .
\ee
From the vector boson masses we find a tree-level $\rho$ parameter
equal to:
\be
\rho={x^2\over x^2+y^2}=1-{s_W^2\over x^2 c_W^2}
\label{trho}
\ee
To suppress the unacceptable contribution to $\Delta\rho$, one should
consider large values of the $x$ parameter. In turn, a large $g_R$ 
coupling constant does not necessarily imply large observable effects 
on the ordinary particles, since the mixing angles scale as $1/x$.

By analysing the explicit expression of the one-loop contribution 
to $\Delta\rho$ from the scalar sector, we have also found other 
solutions giving a one-loop vanishing result. In particular, we 
would like to mention a variant of the solution discussed above, 
corresponding to the choices: $v_1=v_2=v$, $v_R=v_L/x$ and $\tan
\beta_\pm=\tan\beta_0=-m_{W'}/m_W$. In this case, we no longer 
require $v \gg v_R$, but we fix the mixing angles in the scalar sector 
to a particular non-vanishing value. We obtain:
\bea
\Delta\rho&=&{x^2\over 16 \pi^2 v_L^2 (1+x^2)}
\left\{2({m_1^\pm}^2 
+m_5^2)+4 g(m_{5},m_{1}^{\pm}) \right.\nn\\ 
& &\left.+(1-{1\over \tan\beta^2})^2 [{m_2^\pm}^2 - g(m_2^0,m_6)+
g (m_{2}^{\pm},m_{6}) + g (m_{2}^{\pm},m_{2}^{0})]\right.\nn\\ 
& &\left. +{1\over \tan\beta^2}[2({m_2^\pm}^2 
+m_2^{0 \, 2})+4 g(m_2^0,m_{2}^{\pm})]\right.\nn\\
& &\left.+({1\over \tan\beta^2}-{1\over \tan\beta^4}) 
(g(m_2^\pm,m_4)-g(m_2^0,m_4))
\right\}
\eea
This contribution vanishes when $m_1^\pm=m_5$ and $m_2^0=m_2^\pm$.
The tree-level $\rho$ parameter is still given by eq. (\ref{trho})
and agreement with data requires large $x$ values. 
To establish an allowed range for $x$ we have fitted again the
electroweak data of table 3 in this special case. The two defining
conditions reduce to two the number of independent parameters,
that we have chosen to be $x$ and $m_{W'}$. As before we have fixed 
$m_t=175 \gev$, $\alpha_s(m_Z)=0.118$ and we have considered
the two cases $m_h=100 \gev$ and $m_h=300 \gev$. We have found
no sensitivity of the fit to the $m_{W'}$ parameter, which has been 
kept fixed to several values in the range $(100,1000)~{\rm GeV}$.
We found no significant improvement with respect to the SM case, 
recovered in the large $x$ limit, and $x > 16$ at the 2 $\sigma$ level.

Notice that in the present case a large $x$ does not necessarily 
mean large $W'$ and $Z'$ masses. From eq. (\ref{mas2}) we see 
that a large $x$ can be compensated by a small $v$. Indeed,
this is the only case we found where a one-loop vanishing 
contribution to $\Delta\rho$ from the scalar sector can be
compatible with relatively light new vector bosons.

We have checked that, in all configurations of VEVs described 
above where the quadratic scalar contribution vanishes, also 
the contribution quadratic in the mass of the new gauge bosons 
$W^{\prime}$, $Z^{\prime}$ cancels. The cancellation always 
occurs inside each individual self-energy ($\Sigma_{WW},\Sigma_{ZZ},
\Sigma_{\gamma Z}$), and involves not only diagrams with gauge 
particles running in the loop, but also those with gauge and 
scalar internal lines (only $G^{\prime}$). The remaining 
contribution is at most logarithmic in the $W^{\prime},
Z^{\prime}$ masses. 

It is interesting to note that, when $v_1=v_2=v$, $v_R=v_L/x$,
it is possible to find contact with the so-called BESS model
\cite{bess}, which couples a triplet of new vector bosons
to the SM particles by means of a non-renormalizable, effective
lagrangian. The model, designed to describe general features
of schemes of dynamical breaking of the electroweak symmetry,
possesses no physical scalar particle. The relation of BESS
to the FP model should then be looked for
in the gauge boson and fermion sectors. BESS is described,
in its minimal form, by 5 parameters (see ref. \cite{bess}): a VEV $f$,
three gauge coupling constants $g$, $g'$ and $g''$ and a 
dimensionless coupling $\alpha$. The particular case we are dealing
with is also characterized by 5 parameters, due to the 2 conditions
among the VEVs imposed to screen $\Delta\rho$ from large one-loop
corrections. We can take $g$, $x$, $y$, $v$ and $v_R$ as free parameters.
It turns out that generic values of these parameters do not reproduce
the relations of the BESS model. It is however sufficient to
require the additional condition $v_R=\sqrt{2} v$ in order to
recover the same results of BESS for the vector boson masses,
the mixing angles and the fermionic interaction terms.
On its side, the BESS model is constrained by the additional
relation $\alpha=2 g^2/(2 g^2+g''^2)$, which restricts to 4
the number of independent parameters. For the interested reader,
we collect in table~\ref{besstab} the dictionary from the BESS model
to the present one, both subject to the supplementary conditions
needed to relate the models.
\begin{table}[htb]
\label{besstab}
\begin{center}
\begin{tabular}{|c|c|}
\hline
BESS model & Present model \\
\hline
$\alpha$ & ${1\over (1+2 x^2)}$\\
\hline
$f$ & $v\sqrt{1+2 x^2}$\\
\hline
$g''$ & $2 g x$\\
\hline
$g$ & $g$\\
$g'$ & $\tilde{g}$\\
\hline
\end{tabular}
\end{center}
\caption{{ \it Translation table between the BESS model, subject to
the condition $\alpha=2 g^2/(2 g^2+g''^2)$ and the present model,
subject to the conditions $v_1=v_2=v$, $v_R=v_L/x=\sqrt{2} v$.}}
\end{table}
By relaxing the condition $v_R=v_L/x$, it is possible to find a 
one-to-one correspondence between the two models in the full 5-fold
parameter space, as already noticed in the last of refs.~\cite{fp}. 
In this case, however, the one-loop contributions to $\Delta\rho$ 
from the scalar sector are quadratic in the scalar masses and hardly 
reconcilable with the data.

\subsubsection{$R_b$}

At the classical level, the SM prediction for $R_b$ can
be modified by mixing effects in the gauge boson sector,
as accounted for in our tree-level fit to electroweak 
precision data. We have
explicitly verified that, in the region of parameter 
space allowed by our fit, the shift in the predicted 
value of $R_b$ with respect to its SM value is always
negligible, i.e. smaller in absolute value than $10^{-4}$.
This suggests that non-negligible (and 
non-SM) loop corrections to $R_b$ may possibly come
only from loops involving the extra scalar particles
of the FP model. Loops involving the exchange of neutral
Higgs bosons are controlled by couplings proportional
to the $b$-quark mass, and cannot give large effects.
Similarly, the one-loop $W'$ contribution is suppressed
by $(\alpha_\pm)^2$.
The only contributions that deserve a more accurate
study are those associated with the exchange of 
virtual charged Higgs bosons. For simplicity, we 
work in the limiting case of eqs.~(\ref{limcase}) 
and (\ref{chyukbis}). We can then express the additional 
contribution to $R_b$ due to charged Higgs exchange as
\cite{rbhiggs}
\be
\label{deltarb}
\Delta R_b \simeq (R_b)_{SM} \times 0.78 \times
{\alpha_W \over 2 \pi} {v_L  \over v_L^2 + v_R^2}
\times F_H  \, ,
\ee
where $(R_b)_{SM} \simeq 0.2158$ is the SM prediction,
$\alpha_W \equiv 4 \pi g^2$ and
\be
v_L = - {1 \over 2} + {1 \over 3} \sin^2 \theta_W \, ,
\;\;\;\;\;
v_R = {1 \over 3} \sin^2 \theta_W \, .
\ee
The function $F_H$, associated with the top-Higgs loops,
depends on the common mass $m_H$ of the charged Higgs
bosons and on their dominant coupling to top and bottom
quarks, eq. (\ref{chyukbis}):
\be
\label{lambdah}
\lambda_H = {m_t \over \sqrt{2} m_W} {\tan \theta_W
\over \sqrt{x^2 - \tan^2 \theta_W}} \, ,
\ee
and reads
\be
\begin{array}{c}
F_H = \left\{ b_1(m_H,m_t) v_L - c_0 (m_t,m_H) v_L^{(H)}
+ m_t^2 c_2(m_H,m_t) v_L^{(t)} \right.
\\ \\
+ \left.
\left[ m_Z^2 c_6(m_H,m_t) - {1 \over 2}
- c_0(m_H,m_t) \right] v_R^{(t)} \right\}
\lambda_H^2 \, .
\end{array}
\ee
The functions $b_1$, $c_0$, $c_2$ and $c_6$ can be
found, for example, in the appendix of ref.~\cite{bfz}, and
\be
v^{(t)}_L = {1 \over 2} - {2 \over 3} \sin^2 \theta_W \, ,
\;\;\;\;\;
v^{(t)}_R = - {2 \over 3} \sin^2 \theta_W \, ,
\;\;\;\;\;
v^{(H)}_L = - {1 \over 2} + \sin^2 \theta_W \, .
\ee
\begin{figure}[ht]
\vspace{-0.1cm}
\epsfig{figure=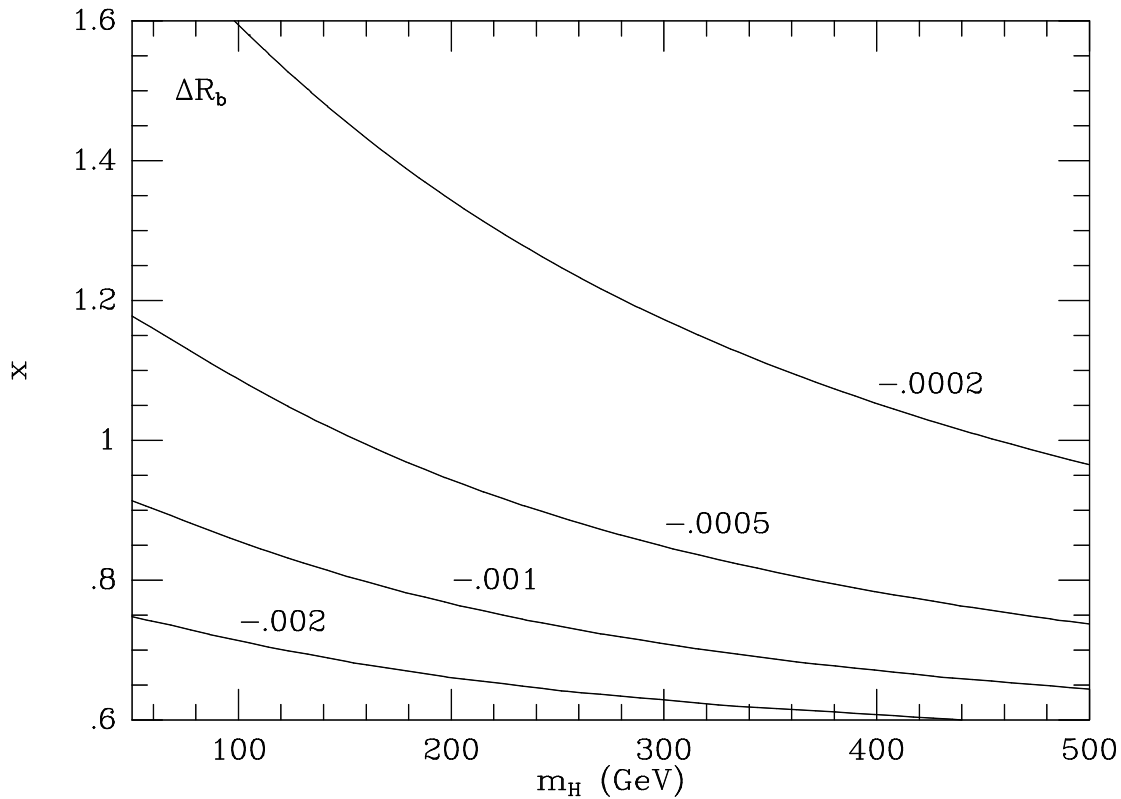,height=6cm,angle=-90}
\vspace*{3.5cm}
\caption{{\it Contours of $\Delta R_b$ in the
($m_H,x$) plane.}}
\label{figrb}
\end{figure}
Fixing $m_t = 175$ GeV, we have explored the possible values 
of $\Delta R_b$ as functions of $m_H$ and $x$, as shown by the 
contours in fig.~\ref{figrb}. It is important to notice that in 
the case under consideration $\Delta R_b$ is always negative,
and sizeable effects can be obtained for small $m_H$ and
$x$ close to $\tan \theta_W$, corresponding to a strong $\lambda_H$
coupling. Since the world average for $R_b$, given in table~3,
is $1.75 \sigma$ in excess with respect to the SM prediction,
with an experimental situation still in rapid evolution, we choose
$\Delta R_b > - 0.001$ as a tentative bound. We can then see that
a significant region of the $(m_H,x)$ plane can already be excluded.

\subsection{Contributions to flavour-changing processes }

In general, models based on the gauge group $SU(2)_L \times SU(2)_R 
\times U(1)_{\tilde{Y}}$ are very strongly constrained by 
experimental data from flavour physics, in particular by FCNC 
processes \cite{lrflav}. In this respect, the FP model has a privileged 
status, since it automatically guarantees the absence of FCNC
at tree-level and the suppression of loop-induced effects, 
thanks to the fact that the unmixed $SU(2)_R$ gauge bosons and 
the $(\phi_{LR},\phi_R)$ Higgs bosons do not have direct couplings 
to the matter fermions. 
 
Choosing $|\alpha_\pm| < 0.01$, as suggested by our fit to 
electroweak observables, and $m_{W^\prime} \simgt m_W$, 
we can estimate a negligible one-loop 
contribution to the relevant observables from $W'$ exchange.
In the present discussion such contribution can be safely 
omitted, and our focus will be on the charged Higgs boson 
exchanges that, for flavour-changing phenomena, dominate the 
one-loop corrections of non-standard origin whenever they are 
non-negligible. To gauge the typical effects from the charged 
scalar sector we will work in the limit of 
eqs.~(\ref{limcase}--\ref{chyukbis}).

Before moving to loop-induced FCNC processes, it is useful to
review the limits on the charged Higgs sector that come from
tree-level charged-current processes, such as heavy flavour
decays. The process $b \to c \tau \nu_{\tau}$, that originates
non-trivial constraints in other multi-Higgs models \cite{bctau}, 
is of no use in the FP model, since the Yukawa couplings proportional
to the $b$ and $\tau$ masses are always much smaller than those
proportional to the $t$ mass. This is an obvious consequence
of the fact that in the FP model only $\phi_L$ is coupled to 
fermions.

Interesting limits can instead be obtained by considering the
decay $t \to b H^+$, which competes with the SM channel $t \to b 
W^+$. In the limiting case of eqs.~(\ref{limcase}) and (\ref{chyukbis}), 
the partial widths for $t \to b H^+$ and $t \to b W^+$ read:
\be
\Gamma (t \to b H^+ ) =
{\sqrt{
[m_t^2 - (m_H + m_b)^2]
[m_t^2 - (m_H - m_b)^2]}
\over 16 \pi m_t^3} \cdot {\cal A}_H \, ,
\label{gammatbh}
\ee
\be
\label{aacca}
{\cal A}_H  = {g^2 \over 4 m_W^2}
{\tan^2 \theta_W \over x^2 - \tan^2 \theta_W}
\left[ ( m_t^2 + m_b^2 -m_H^2) (m_b^2 + m_t^2) - 4 m_b^2 m_t^2
\right] \, ;
\ee
\be
\label{gammatbw}
\Gamma (t \to b W^+ ) =
{\sqrt{
[m_t^2 - (m_W + m_b)^2]
[m_t^2 - (m_W - m_b)^2]}
\over 16 \pi m_t^3} \cdot {\cal A}_W \, ,
\ee
\be
\label{avdoppio}
{\cal A}_W  = {g^2 \over 4 m_W^2} \left[
m_W^2 (m_t^2 + m_b^2 - 2 m_W^2) + (m_t^2 - m_b^2)^2 
\right] \, .
\ee
\begin{figure}[ht]
\vspace{-0.1cm}
\epsfig{figure=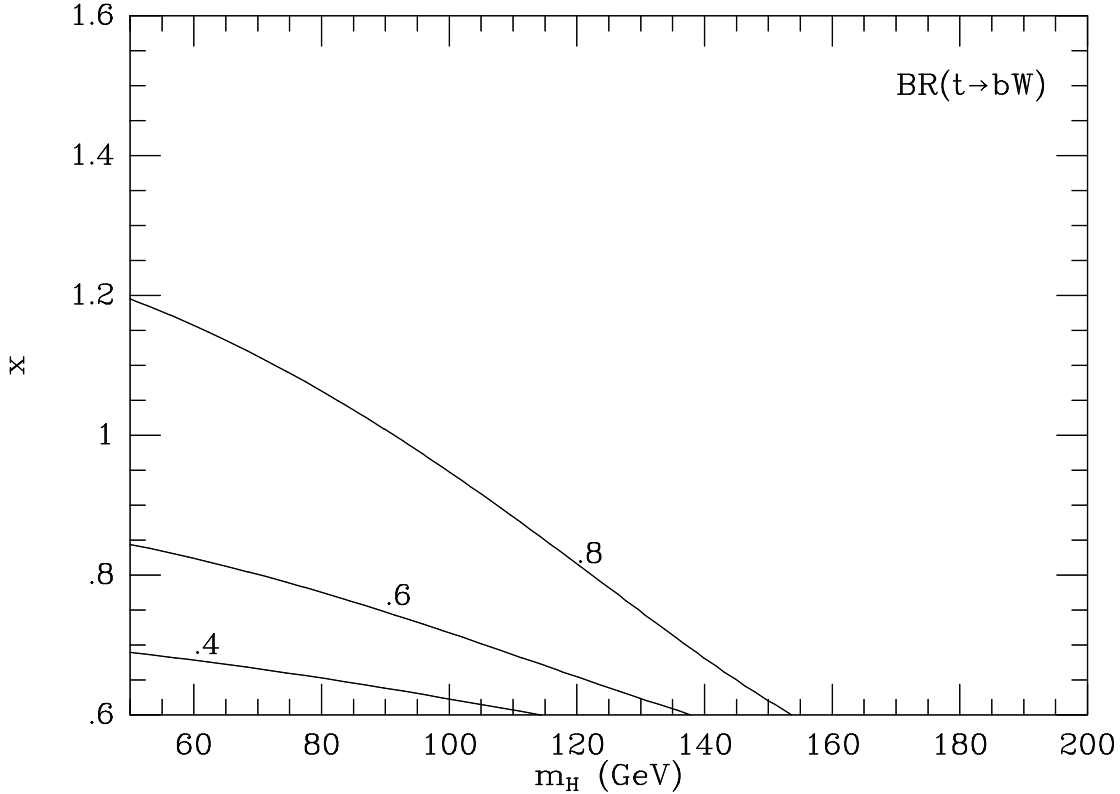,height=6cm,angle=-90}
\vspace*{3.5cm}
\caption{{\it Contours of $BR(t \to b W)$ in the
($m_H,x$) plane.}}
\label{figtbh}
\end{figure}
With the help of fig.~\ref{figtbh}, which displays contours of 
$BR(t \to b W^+)$ in the $(m_H,x)$ plane, we can see that deviations 
from the SM prediction $BR(t \to b W^+) \simeq 1$ can be very significant,
up to $BR(t \to b W^+) \sim 0.4$. However, this requires some work
to be transformed into a constraint on the parameter space, since
the Tevatron experiments use to give their bounds on charged Higgs 
bosons \cite{tbhbounds} in terms of the parameters $(\tan \beta,m_H)$,
as defined in a special subclass of two-doublet models, and in any case 
these bounds have some dependence on the assumed top production 
cross-section. As a tentative reference value for the CDF and D0 
sensitivity, we can take $BR (t \rightarrow b W^+) = 0.6$. Even this 
conservative estimate is sufficient to rule out a significant region 
of the $(m_H,x)$ plane, characterized by low values of $m_H$ and $x$.

\subsubsection{ $ b \to s \gamma $}

The experimental determination \cite{cleo} of the inclusive $B 
\to X_s \gamma$ branching ratio, $BR(B \to X_s \gamma) = (2.32 
\pm 0.67) \times 10^{-4}$, has been recently supplemented by 
the complete next-to-leading-order SM calculation \cite{bsgth},
giving $BR(B \to X_s \gamma)_{SM} = (3.28 \pm 0.33) \times 10^{-4}$.
These two results strongly constrain many possible extensions 
of the SM, and in particular the FP model, as we shall now see. 
\begin{figure}[htb]
\vspace{-0.1cm}
\epsfig{figure=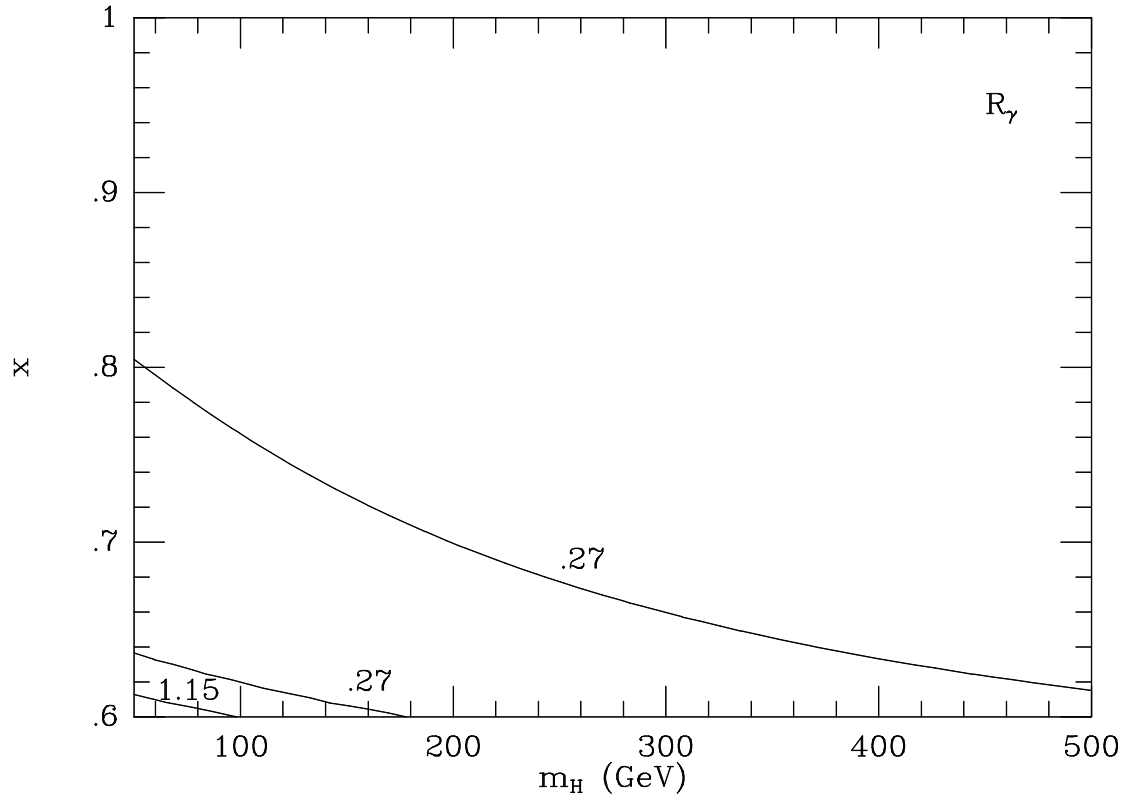,height=6cm,angle=-90}
\vspace*{3.5cm}
\caption{{\it Contours of $R_{\gamma}$ in the ($m_H,x$) plane.}}
\label{figrg}
\end{figure}
Both in the SM and in the FP model, the dominant contribution
comes from the effective operator ${\cal O}_7 \propto m_b 
\ov{s_L} \sigma^{\mu\nu} b_R F_{\mu \nu}$. Following the strategy 
of \cite{bfz}, we express our results in terms of the ratio
\be
\label{rgamma}
R_{\gamma} \equiv
\frac{Br (B \to X_s \gamma)_{FP}}{Br(B \to X_s \gamma)_{SM}}
\simeq
\left[ \frac{ C (A_W + A_H)_{FP} + D }
 { C~ (A_W)_{SM} + D } \right]^2 \, ,
\ee
where $C \simeq 0.66$ and $D \simeq 0.35$ take into account the 
leading QCD corrections. In the SM, the dominant one-loop 
diagrams involve the exchange of virtual $W$ bosons and top quarks,
and give
\be
(A_W)_{SM} =  
x_{t W} \left[ 2 \, F_1(x_{t W}) + 3 \, F_2(x_{t W}) \right] \, . 
\label{eq:wampl}
\ee
Here and in the following, we set by convention $x_{i j} = m^2_i / 
m^2_j$. The explicit expression of the different $F$-functions can 
be found in the appendix of ref.~\cite{bfz}. In the FP model, 
considering the limit of vanishing mixing angle $\alpha_\pm$, 
gauge boson exchange gives the same result as in the SM,
eq.~(\ref{eq:wampl}).
One should add to the previous result the contributions from
one-loop diagrams involving the exchange of virtual top
quarks and charged Higgs bosons, which can be significant \cite{bsghiggs}.
Working as before in the limiting case of eq.~(\ref{limcase}):
\bea
A^{FP}_H = 
\dd\frac{x_{t H}}{3}\dd\frac{t_W^2}{x^2-t_W^2} 
\left[ 2 F_1(x_{t H}) + 3 F_2(x_{t H}) -
2 F_3(x_{t H}) - 3 F_4(x_{t H}) \right] \, .
\eea
The possible values of $R_{\gamma}$ in the $(m_H,x)$ plane are 
shown by the contour plot of fig.~\ref{figrg}. Our conservative 
estimate of the presently allowed range of variation is
\be
\label{rgbound}
0.27 < R_{\gamma} < 1.15 \, .
\ee
We then see that the constraint of eq.~(\ref{rgbound}) excludes 
simultaneously small values of $m_H$ and $x$, apart from a small 
strip near $x=0.6$ and $m_H = 100 \gev$, which is however excluded 
by other constraints. Notice that, in contrast with other popular
models, in the FP model a light charged Higgs is likely to give
$R_{\gamma} < 1$.

\subsubsection{$B_0 - \bar B_0$ and $K_0 - \bar K_0$ mixing } 
\label{sec:bb}

We discuss here the FP-model contributions to the 
$B_d^0$--$\ov{B_d^0}$ mass difference $\Delta m_{B_d}$ 
and to the CP-violation parameter of the $K^0$-$\ov{K^0}$ 
system $\epsilon_K$, and the constraints on the model 
parameters coming from the experimentally measured
values of $\Delta m_{B_d}$ and $\epsilon_K$. As usual,
we consistently neglect terms proportional to $m_b$
in the charged Higgs vertices with top and bottom quarks. 

For our purposes, a convenient way of parametrizing 
the $B_d^0$--$\ov{B_d^0}$ mass difference is:
\be
\Delta m_{B_d} = \eta_{B_d} \cdot {4 \over 3}
f_{B_d}^2 B_{B_d} \cdot m_{B_d} \cdot
\left( \alpha_W \over 4 m_W \right)^2
\cdot \left| K_{tb} K^*_{td} \right|^2
\cdot x_{tW} \cdot | \Delta | \, ,
\ee
where $\eta_{B_d} \simeq 0.55$ is a QCD correction factor;
$f_{B_d}$ is the $B_d$ decay constant and $B_{B_d}$ the vacuum
saturation parameter. The quantity $\Delta$ contains the 
dependence on the parameters of the FP model. We have checked 
that, for values of $\alpha_\pm$ and $m_{W'}$ allowed by other
constraints, the contributions to $\Delta$ coming from
box diagrams with internal $W'$ lines can be safely 
neglected. We can then perform the following decomposition:
\be
\label{dm}
\Delta = \Delta_W + \Delta_H  \, .
\ee
In eq.~(\ref{dm}), $\Delta_W$ denotes the Standard Model
contribution, associated with the box diagrams involving
the top quark and the $W$ boson:
\be
\Delta_W =   A ( x_{tW} ) \, ,
\ee
where the explicit expression of the function $A(x)$ can
be found in the appendix of ref.~\cite{bfz}. $\Delta_H$ 
denotes the additional
contributions from the box diagrams involving the physical
charged Higgs bosons \cite{buras}. Working as before in
the limiting case of eq.~(\ref{limcase}), we find:
\be
\Delta_H = \lambda_H^4 \, x_{Wt} \, x_{WH} G(x_{tH})
+  \lambda_H^2 \left[ 4 F \, '(x_{tW}, x_{HW})
+  G \, ' (x_{tW} , x_{HW}) \right] \, ,
\ee
where $\lambda_H$ has been defined in eq.~(\ref{lambdah}), and 
the functions $G(x)$, $F \, ' (x,y)$ and $G \, ' (x,y)$  are 
given in the appendix of ref.~\cite{bfz}. 

Moving to the $K^0$--$\ov{K^0}$ system, the absolute value of the
parameter $\epsilon_K$ is well approximated by the expression:
\be
\vert \epsilon_K\vert=
{2 \over 3} f_{K}^2 B_{K} \cdot
\frac{m_K}{\sqrt{2} \Delta m_K} \cdot
\left( \alpha_W \over 4 m_W \right)^2
\cdot x_{cW} \cdot
| \Omega | \, ,
\label{epsk}
\ee
where $f_K$ is the $K$ decay constant, $B_K$ is the vacuum
saturation parameter (recently re-evaluated in \cite{ioioio}), 
$\Delta m_K$ is the experimental $K^0_L$--$K^0_S$ mass difference. 
The quantity $\Omega$, carrying the dependence on the mixing angles 
and the FP-model parameters, is given by:
\be
\Omega=\eta_{cc}~ {\rm Im} (K_{cs} K_{cd}^*)^2  +
2 \eta_{ct}~ {\rm Im} (K_{cs} K_{cd}^* K_{ts} K_{td}^*)^2~
[ B(x_{tW})- \log x_{cW} ] +
\eta_{tt}~ {\rm Im} (K_{ts} K_{td}^*)^2~ x_{tc}~\Delta \, ,
\label{omega}
\ee
where $\eta_{cc} \simeq 1.38$, $\eta_{ct} \simeq 0.47$ and
$\eta_{tt} \simeq 0.57$ are QCD correction factors; $x_{cW}
= m_c^2 / m_W^2$, $x_{tc} = m_t^2/m_c^2$; the function
$B(x)$ can be found in the appendix of ref.~\cite{bfz}; 
$\Delta$ is the same as in
eq.~(\ref{dm}). In principle, there are additional contributions 
due to charged Higgs exchange besides those appearing in $\Delta$.
However, in the FP model they can be safely neglected with
respect either to the standard contribution or to the 
non-standard contribution parametrized by $\Delta$, hence 
they have not been considered here.

\begin{figure}[htb]
\vspace{-0.1cm}
\epsfig{figure=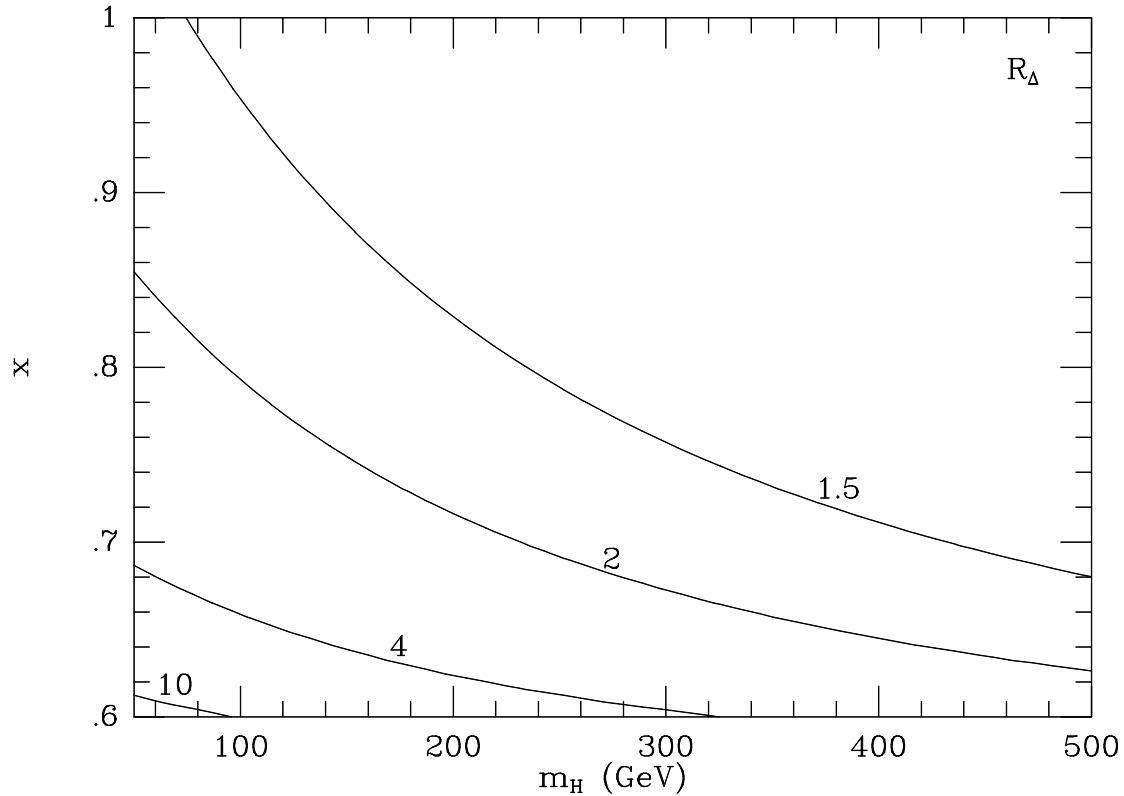,height=6cm,angle=-90}
\vspace*{3.5cm}
\caption{{\it Contours of $R_{\Delta}$ in the ($m_H,x$) plane.}}
\label{figrd}
\end{figure}

We have studied the dependence of $\Delta$ on the parameters 
$(m_H,x)$, characterizing the charged Higgs sector. We observe
that $\Delta_{FP} > \Delta_{SM}$. Some quantitative information 
is given in fig.~\ref{figrd}, which displays contours of the
ratio
\be
\label{rdelta}
R_{\Delta} \equiv {\Delta \over \Delta_W}
\ee
in the plane $(m_H,x)$. Observe that values of $R_{\Delta}$ 
much larger than 1 can be obtained for small values of $m_H$ 
and of $x$. 

To discuss the constraints coming from the measured values of
$\Delta m_{B_d}$ and $\epsilon_K$, we recall that the dependence 
on the FP-model parameters is contained in the quantity $\Delta$ 
of eq.~(\ref{dm}), so it would be desirable to obtain from the 
experimental data a bound on $\Delta$. On the other hand, this 
requires some knowledge
of the parameters characterizing the mixing matrix $K$. Notice that
we cannot rely upon the SM fit to the matrix $K$, since among the
experimental quantities entering this fit there are precisely $\Delta
m_{B_d}$ and $\epsilon_K$, whose description now differs from the SM
one.

To derive the desired bound on $\Delta$, we have used the results
of the fit performed in \cite{bfz}. As discussed there, it is not 
straightforward to translate those results into a
single definite bound on $\Delta$, or, equivalently, on $R_{\Delta}
= \Delta / \Delta_W \simeq 1.8 \Delta$. As a tentative bound
we can consider here $0.4 < R_{\Delta} < 4$. Contours of $R_{\Delta}$
in the $(m_H,x)$ plane are shown in fig.~\ref{figrd}: we can see
that small values of $m_H$ and $x$ are excluded.
\subsection{$W'$ and $Z'$ signals at hadron colliders}

In this section we analyse possible signals of the new  
vector bosons of the FP model at the Tevatron collider and at 
the LHC. We will obtain new restrictions on the parameter space 
of the FP model and describe some of its specific signatures. 
From the previous sections, we know that relatively light $W'$ 
and $Z'$ are not excluded, provided that $x$ is sufficiently
large. The neutral gauge boson $Z'$ possesses a direct coupling
to ordinary fermions that scales as $1/x$, and also an indirect 
coupling, via the mixing controlled by the angle $\alpha_0$,
subject to the phenomenological restriction $|\alpha_0|
\simlt 10^{-2}$. On the other hand, the charged vector boson 
$W'$ can couple to fermions only through the mixing controlled 
by the angle $\alpha_\pm$, also subject to a similar 
phenomenological bound, $|\alpha_{\pm}| \simlt 10^{-2}$.
 
Both $Z'$ and $W'$ can be produced at hadron colliders via
quark-antiquark annihilation. In tables~7 and~8 we show
the total cross-sections for the production of $Z'$ and $W'$,
respectively, at the Tevatron collider, $\sqrt{s}=1.8~{\rm TeV}$.
The cross-sections have been evaluated in the narrow width 
approximation, using the parton densities of \cite{tung}. A 
K-factor $\simeq 1.2$ has been included.

The cross-sections of table~7 were computed in the limit 
$\alpha_0=0$. We checked that only small variations are induced
by varying the mixing angle in the range $|\alpha_0|\le 10^{-2}$.
Indeed, we expect a dependence on $\alpha_0$ only for large $x$,
when the direct coupling and mixing effects become comparable.
For $|\alpha_0| \sim 10^{-2}$, such a dependence would manifest 
approximately at $x \sim 10^2$, beyond the range explored here.
Notice that the $Z'$ cross section scales approximately as $1/x^2$, 
as expected from the $x$ dependence of its couplings to fermions.
\begin{table}[htb]
\begin{center}
\begin{tabular}{|c|c|c|c|c|}
\hline
$m_{Z'}$ (GeV)& $x=0.6$ & $x=1$ & $x=5$ & $x=20$
\\
\hline
100 & $1.4\cdot 10^4$ & $12.5\cdot 10^2$ & $35.4$ & $2.2$
\\
\hline
250 & $8.1\cdot 10^2$ & $71.0$ & $2.0$ & $12.5\cdot 10^{-2}$
\\
\hline
500 & $41.2$ & $3.6$ & $10.2\cdot 10^{-2}$ & $63.2\cdot 10^{-4}$
\\
\hline
1000 & $9.3\cdot 10^{-2}$ & $81.0\cdot 10^{-4}$ & $2.3\cdot 10^{-4}$ & 
$14.2\cdot 10^{-6}$
\\
\hline
\end{tabular}
\end{center}
\caption{{\it Total cross-section in $pb$ for $Z'$ production at 
the Tevatron collider, $\sqrt{s}=1.8~{\rm TeV}$ for $\alpha_0=0$.}}
\end{table}
\begin{table}[htb]
\begin{center}
\begin{tabular}{|c|c|c|c|c|}
\hline
$m_{W'}$ (GeV)& $100$ & $250$ & $500$ & $1000$
\\
\hline
$\sigma_{W'}(pb)$ & $0.6$ & $2.8\cdot 10^{-2}$ & $9.8
\cdot 10^{-4}$ & $1.3\cdot 10^{-6}$
\\
\hline
\end{tabular}
\end{center}
\caption{{ \it Total cross-section in $pb$ for $W'$ production at the 
Tevatron collider, $\sqrt{s}=1.8~{\rm TeV}$ for $\alpha_\pm=10^{-2}$.}}
\end{table}

The $W'$ cross-section scales as $(\alpha_\pm)^2$. Moreover, it is
independent of $x$, since $W'$ is coupled to the standard $SU(2)_L$ 
current. In table~8 we considered $\alpha_\pm=0.01$, at the border 
of the region allowed by precision tests. Even in this case the 
$W'$ cross-section is quite modest, below the observability level as 
soon as $m_{W'}$ is larger than 250 GeV. The big difference between 
the $Z'$ and $W'$ cross-sections listed in tables~7 and~8 is largely 
due to the different interaction properties of $Z'$ and $W'$ with 
fermions. Due to the suppression of $W'$ production at hadron colliders, 
the only significant limitations on the parameter space from the 
Tevatron data are those that can be obtained through the study of 
the $Z'$ channel.

In the range of parameters considered in table~7,  the $Z'$ 
cross-section at the Tevatron collider is sizeable and might 
have produced an observable signal. Beyond the traditional 
dilepton channel \cite{tevll}, the CDF and D0 collaborations
have recently searched for $Z'$ in the dijet and in the $b {\bar b}$
channels \cite{tevjj}. Moreover, the same collaborations have 
measured the cross-sections for diboson production \cite{tevvv}, 
which can be modified in the presence of a $Z'$. Indeed, the $Z'$ 
of the FP model can decay into fermion-antifermion pairs, or in $WW$, 
$WW'$, $W'W'$ and $Z h$, when kinematically possible. 

In the limit of vanishing mixing angles $\alpha_0$ and $\alpha_\pm$, 
the tree-level interaction terms $Z'WW$, $Z'WW'$, $Z'Zh$
vanish together with the corresponding $Z'$ partial widths. 
In this approximation $Z'$ decays almost exclusively in leptons or
quark pairs, in the ratios 15:3:5:17 for (massless) $e^+ e^-$, 
$\nu {\bar\nu}$, $u{\bar u}$ and $d{\bar d}$, respectively. On the 
experimental side, the sensitivity is larger for the dilepton channel
($e^+ e^-$ and $\mu^+ \mu^-$) than for the dijet or $b{\bar b}$ 
channels. The dilepton search provides the most stringent constraint 
on the FP model. 

For non-vanishing mixing angles, the branching ratios of $Z'$
into $WW$ and $Zh$ can become comparable with those into fermions.    
For instance, it is well known \cite{dqz} that in the $WW$ channel 
the suppression ${\alpha_0}^2$ in the squared coupling constant can be 
compensated by the kinematical factor $(m_{Z'}/m_{W})^4$, for 
sufficiently large $m_{Z'}$. Moreover, the fermionic modes can be 
depleted by a large $x$ value.

\begin{figure}[htb]
\vspace{-0.1cm}
\epsfig{figure=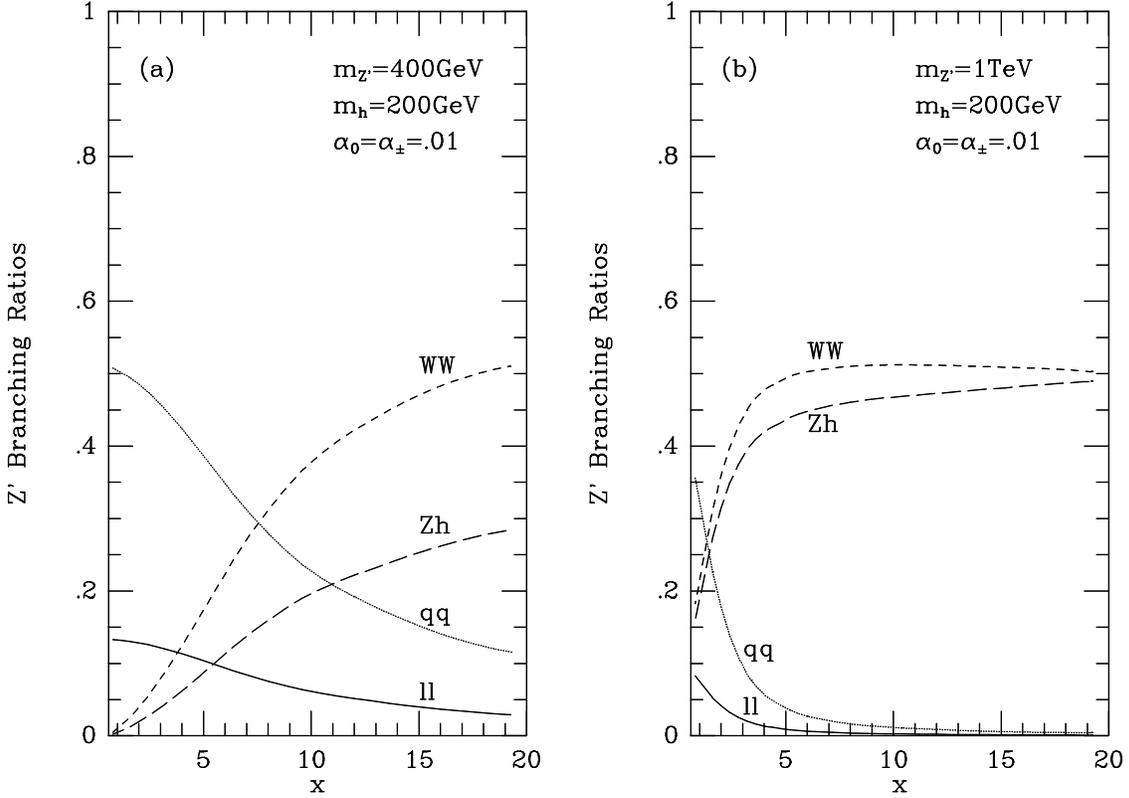,height=6cm,angle=-90}
\vspace*{3.5cm}
\caption{{\it $Z'$ branching ratios, as functions of $x$, for
$m_h=200 \gev$, $\alpha_0 = \alpha_{\pm} = 0.01$ and: (a) $m_{Z'} 
= 400 \gev$; (b) $m_{Z'} = 1 \tev$.}}
\label{figbrzp}
\end{figure}

In fig.~\ref{figbrzp} we show some of the $Z'$ branching ratios, 
as functions of $x$, for $m_{Z'}=400,1000 \gev$, $\alpha_0 =
\alpha_\pm = 0.01$ and $m_h=200~{\rm GeV}$: the line denoted by 
$ll$ corresponds to decays into charged lepton pairs of a single
generation, the one denoted by $qq$ to decays into all possible
quark-antiquark pairs; the branching ratio for the decays into
neutrino pairs is not shown. When $x$ is close to its lower bound, 
the fermionic channels are enhanced due to the large coupling constant. 
Moving to larger values of $x$, the fermionic branching ratios decrease. 
When $m_{Z'} = 400 \gev$, they are reduced by a factor 5 going from $x 
= 0.6$ to $x = 20$. When $m_{Z'} = 1000 \gev$, the reduction factor is 
about 160 in the same $x$ interval. The larger suppression for larger 
values of $m_{Z'}$ is due to the positive powers of $(m_{Z'}/m_{Z})$ 
that characterize the diboson channels. Moreover, when $m_{Z'}$ is 
large, smaller values of $x$ are needed to obtain significant branching
ratios into $WW$ or $Zh$.
  
In practice, however, for those values of $m_{Z'}$ and $x$ that make 
the diboson channels dominant, the total cross-section for $Z'$ 
production becomes small. We have explicitly verified that the most 
stringent bound from the Tevatron is always the one related to
dilepton searches. 

The total $Z'$ width, $\Gamma_{Z'}$, strongly depends on $x$. 
When $x$ is close to its lower bound, $\Gamma_{Z'}/m_{Z'}$ is 
dominated by the fermionic channels. For $x=0.6$, $\Gamma_{Z'}
/ m_{Z'}$ ranges between 0.07 and 0.085 for $m_{Z'}$ in the interval 
$(100,1000)~{\rm GeV}$. For large $x$ values, the width is saturated 
by the $WW$ and $Zh$ channels, and $\Gamma_{Z'} / m_{Z'}$ never 
exceeds few per mille for $m_{Z'} < 1 \tev$. 

\begin{figure}[htb]
\vspace{-0.1cm}
\epsfig{figure=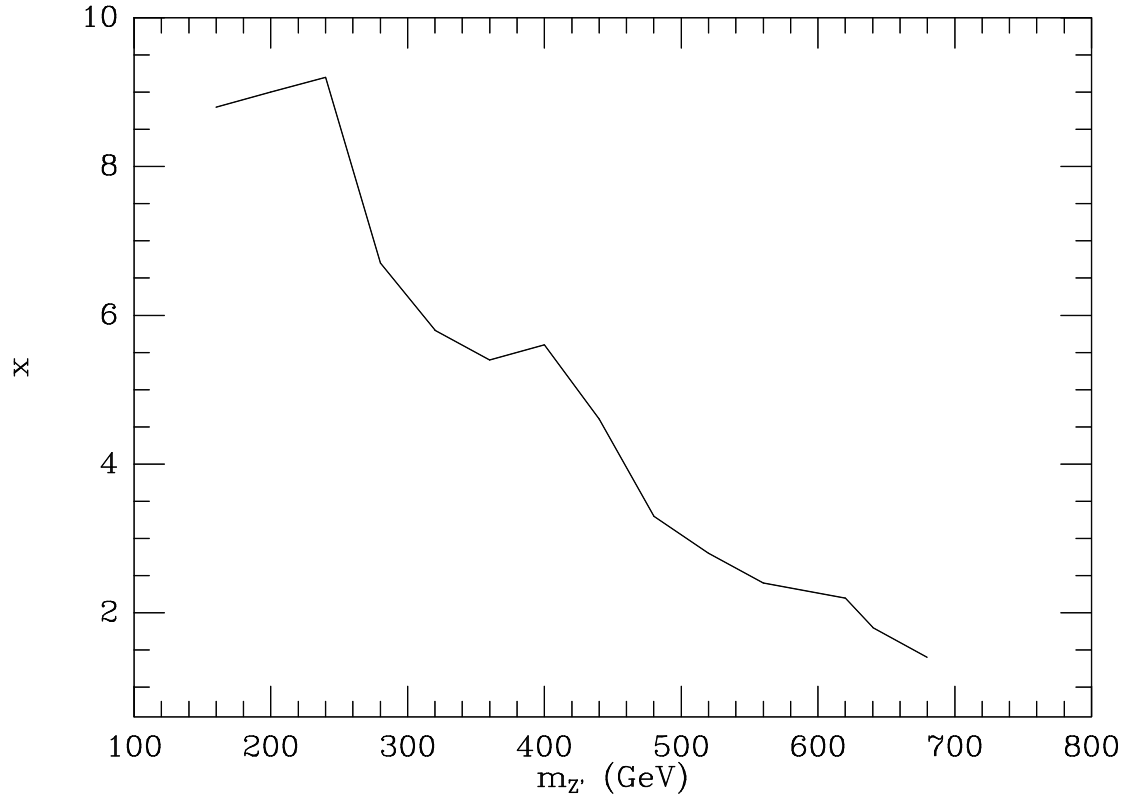,height=6cm,angle=-90}
\vspace*{3.5cm}
\caption{{\it Region of the $(m_{Z'},x)$ plane excluded by 
dilepton searches at the Tevatron collider.}}
\label{figexczp}
\end{figure}

In fig. \ref{figexczp} we present the region of the $(m_{Z'},x)$
plane excluded by dilepton searches 
\footnote{In view of the difficulty of combining the CDF and D0 data
\cite{tevll}, we have tentatively taken a rough interpolation
of the CDF exclusion contour in the $(m_{Z'},\sigma\cdot BR(Z'\to
{\bar l} l))$ plane.}
at $\sqrt{s} = 1.8~{\rm TeV}$. 
We have considered the case $\alpha_0=0$. By turning $\alpha_0$ on, 
the leptonic branching ratio is modified only in extreme regions 
of the $(m_{Z'},x)$ plane. We checked that the exclusion region 
of fig.~\ref{figexczp} is not sensitive to (small) non-zero values
of $\alpha_0$. When $x=1$, a lower bound on $m_{Z'}$ of approximately 
$670~{\rm GeV}$ is obtained, numerically close to the lower bound on 
a $Z'$ with SM couplings. On the other hand, for larger $x$ lighter 
$Z'$ are still allowed by the Tevatron data. For instance, $m_{Z'}=
300~{\rm GeV}$ is permitted when $x > 6$. 

We conclude our discussion about the Tevatron data by adding some 
comments on the special configuration $v_1=v_2=v$ and $v_R=v_L/x$,
selected in section 3.3.1 by the analysis of the one-loop scalar 
contribution to the $\rho$ parameter. We have seen that in this 
case quite large values of $x$ are required 
to obtain a reasonable fit of the electroweak data. The properties 
of $W'$ and $Z'$ are now similar. They are degenerate in mass. The 
mixing angles $\alpha_0$ and $\alpha_\pm$ scale as $1/x$. Direct 
coupling and mixing effects are comparable for $Z'$, and both are 
suppressed by a $1/x$ factor. The branching ratios of $Z'$ and $W'$ 
are now dominated by the fermionic channels. In particular the 
$Z'$ branching ratio into electrons and muons is about $9\%$. For 
$x \simgt 20$ the whole mass range $m_{Z'} > 400~{\rm GeV}$ 
is allowed by the Tevatron data.

Finally, we have looked for possible signals of the FP model at
the LHC. In table~9 we show the total cross-section for $Z'$ 
production at a $pp$ collider with $\sqrt{s} = 14 \tev$.
\begin{table}
\centering
\begin{tabular}{|r|r|r|r|r|}
\hline
$m_{Z'} \; (GeV) $ & x=0.6 & x=1 & x=5 & x=20 \\ 
\hline 
 500  & $6.69 \cdot 10^2$   &       58.5        &       1.66        &
      0.10         \\ 
1000  & 52.8  &       4.62        &       0.13        & $0.8\cdot 10^{-2}$\\ 
2500  & 0.66  &       0.06        &       $0.2\cdot 10^{-2}$ & 
$1 \cdot 10^{-4}$ \\ 
5000  & $0.3\cdot 10^{-2}$ & $2 \cdot 10^{-4}$ & $7 \cdot 10^{-6}$ & 
$4 \cdot 10^{-7}$ \\ \hline
\end{tabular}
\caption{{\it Total $Z'$ cross-section, in $pb$, at the LHC, 
$\sqrt{s}=14~{\rm TeV}$, for $\alpha_0=0$.}}
\label{tab:lhc1}
\end{table}
As for the Tevatron collider, we find only modest variations of the
cross-section when varying the mixing angle $\alpha_0$ in the range 
allowed by the present bounds. Assuming an integrated luminosity
of $10^5~{\rm pb}^{-1}$, from table~\ref{tab:lhc1} we can see that,
at least in principle, even a $Z'$ with $m_{Z'}=5~{\rm TeV}$ 
is within the reach of the LHC, as long as $x < 1$.

We should however pay attention to the $Z'$ branching ratios,
which could vary substantially moving in the allowed parameter 
space. On one side, for very large values of $m_{Z'}$ such as 
those potentially interesting for the LHC, the $WW$ width benefits 
from the huge enhancement factor $(m_{Z'}/m_{W})^4$. On the other 
hand, for fixed values of $x$, $m_{Z'}$ and $\alpha_\pm$, not all 
values of the mixing angle $\alpha_0$ are allowed\footnote{For a 
discussion of the same phenomenon in a different context, see
\cite{dqz}.}. For instance, assuming $\alpha_\pm=0$, the structure
of the neutral gauge boson mass matrix gives rise to the following 
bound on $\alpha_0$:
\be
-\dd\frac{y^2\sqrt{x^2+y^2+x^2 y^2}}{x^2+y^2}
\dd\frac{m_W^2}{(m_{Z'}^2-m_Z^2)}\le
s_0 c_0\le\dd\frac{x^2\sqrt{x^2+y^2+x^2 y^2}}{x^2+y^2}
\dd\frac{m_W^2}{(m_{Z'}^2-m_Z^2)}~,
\label{alphab}
\ee
as can be easily checked by diagonalizing it exactly. For instance,
a mixing angle $\alpha_0=0.01$, allowed by the precision tests,
is incompatible with the simultaneous choices $x=0.6$ and $m_{Z'}=
5 \tev$. Since the bounds in eq.~(\ref{alphab}) scale approximately 
as $m_W^2/m_{Z'}^2$ for large $m_{Z'}$, the enhancement factor is 
totally reabsorbed by the factor $(\alpha_0)^2$, and $\Gamma(Z'\to 
WW)/m_{Z'}$ cannot grow arbitrarily. A similar behaviour holds for 
the partial width $\Gamma(Z'\to Zh)$, which, for asymptotically 
large $m_{Z'}$ values, coincides with $\Gamma(Z'\to WW)$.

\begin{figure}[htb]
\vspace{-0.1cm}
\epsfig{figure=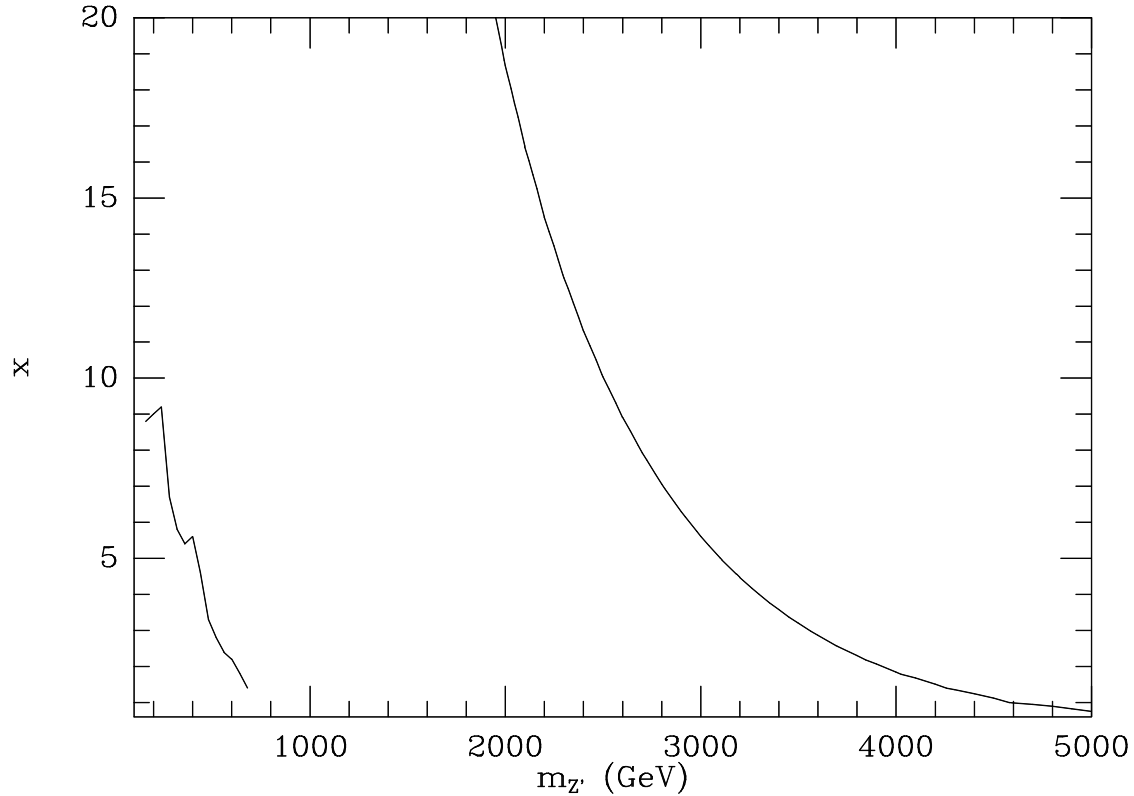,height=6cm,angle=-90}
\vspace*{3.5cm}
\caption{{\it Region of the $(m_{Z'},x)$ plane 
accessible via dilepton searches at the LHC.}}
\label{figlhc}
\end{figure}

A first, rough estimate of the LHC discovery reach can be obtained
by requiring at least 10 events in the $e^+ e^-$ or $\mu^+ \mu^-$ 
channels for an integrated luminosity of $10^5~{\rm pb}^{-1}$, 
neglecting cuts, efficiencies, any detail of the experimental 
apparatus and considering the case of vanishing mixing angles, 
$\alpha_0 = \alpha_{\pm} = 0$, for which the only decay channels are 
the fermionic ones. In fig.~\ref{figlhc} we exhibit the region which 
could be probed by the LHC on the basis of this simple criterium. 
The plot closer to the origin is the exclusion region from the 
Tevatron, presented here for comparison. 

It should be stressed that, when a non vanishing mixing angle $\alpha_0$
is considered, the discovery region of fig. \ref{figlhc} may become smaller,
due to the reduced dilepton branching ratio. Assuming for simplicity a
vanishing mixing in the charged sector, 
the angle $\alpha_0$ can vary in the range defined by eq. (\ref{alphab})
and, depending on the actual values of $(x,m_{Z'})$, the diboson channel
may compete with the dilepton one. For large $m_{Z'}$ and small $x$,
the diboson branching ratios are negligible compared to the leptonic one.
For instance, taking $m_{Z'}=5000~{\rm GeV}$ and $x=0.6$, the fermionic
branching fractions are essentially constant in all the allowed $\alpha_0$ 
range. Indeed, as can be deduced by eq. (\ref{alphab}), such a range becomes
quite narrow for small $x$ and large $m_{Z'}$, and one approaches
the case of vanishing $\alpha$, where the diboson channels are absent.

On the contrary, moving to the region of larger $x$ and smaller $m_{Z'}$,
i.e. climbing up the curve of fig. \ref{figlhc}, the fermionic $Z'$ couplings
decrease and the permitted $\alpha_0$ interval becomes wider, allowing 
for conspicuous diboson branching ratios. For instance, on the points
that correspond to $m_{Z'}= (4,3,2) \tev$ along the LHC contour, we find 
that the dilepton branching ratio can be as small as $(6\cdot 10^{-2},
1\cdot 10^{-3},1\cdot 10^{-4})$, respectively, clearly reducing the
discovery potential of the LHC in the combined ($e^+ e^-, \mu^+ \mu^-)$ 
channel. On the other hand, for the same $m_{Z'}$ values, the $WW$ and 
$Zh$ branching ratios are approximately the same (for light $h$) and are 
given by $(.38,.50,.50)$, respectively. This opens the possibility of
compensating the reduced sensitivity to charged leptons with a dedicated 
search in the diboson channels. In particular, the $Zh$ mode, followed by 
the decay of $Z$ into $l^+ l^-$ $(l=e,\mu)$ and by the decay of $h$ into 
$b {\bar b}$ may provide an interesting signature of the model. The study 
of the corresponding discovery reach at the LHC requires however knowledge 
of acceptances and efficiencies of the experimental apparatus as well as a 
study of the relevant backgrounds within appropriate kinematical cuts, 
which goes beyond the scope of this work.

In conclusion, we have analysed an anomaly-free $SU(2)_R$ extension of 
the SM, with no tree-level FCNC. The $SU(2)_R$ gauge bosons and the 
scalars in the $\phi_R,\phi_{LR}$ multiplets have no coupling with 
the ordinary fermions. Tree-level effects are dominated by a new neutral 
gauge boson $Z'$ that mainly couples to the hypercharge. A comparison 
with the available electroweak data severely constrains the mixing 
angles, both in the neutral and in the charged gauge boson sector, 
still allowing for a region of parameter space with relatively
light new gauge vector bosons. Loop effects may instead be dominated 
by Higgs exchange. We have discussed several possibilities to
cancel the 1-loop contribution to the $\rho$ parameter 
quadratic in the scalar masses.
The remaining loop effects are mainly due to charged Higgs exchange
and are constrained by data on FCNC processes
and by $R_b$. This last constraint is the most restrictive one, and
limits the possible values of the charged Higgs masses and of the $g_R$
coupling. Differently from the usual LR extension of the SM, the
$W'$ contribution to FCNC effects is negligible, thanks to the strict 
limits on the $\alpha_\pm$ angle and to the fermiophobic nature of $W'$.
Also the direct search for a $Z'$ in the dilepton channel at the 
Tevatron collider leads to an excluded region in the $(m_{Z'}, g_R)$ 
plane, which however does not prevent the possibility of relatively 
light new vector bosons, if $g_R$ is sufficiently large. Only the LHC 
collider will have sufficient sensitivity to test new $W'$ and $Z'$ in 
the TeV range.  
\vskip 1.cm

\section*{Acknowledgements}
We would like to thank G.~Chiarelli for useful discussions.
One of us (J.M.) acknowledges financial support from Ministerio
de Educaci\'on y Ciencia (Spain).
\vfill
\newpage
\appendixA{Appendix~A}

We collect here some details of the spectrum and interactions
in the FP model.

The explicit form of the orthogonal $4 \times 4$ matrix $A$,
connecting mass and interaction eigenstates in the charged
Higgs sector and defined by eq.~(\ref{rotcha}), is
\be
A
=\left(
\begin{array}{cccc}
\dd{x v_R\over N_1} t_\alpha &
\dd{x v_R\over N_2} &
e^{(1)}_{1} &
e^{(2)}_{1}
\\
\dd{v_L\over N_1}&
-\dd{v_L\over N_2}t_\alpha &
e^{(1)}_{2} &
e^{(2)}_{2}
\\
\dd{v_1-v_2 x t_\alpha \over N_1}&
-\dd{v_1 t_\alpha +v_2 x\over N_2} &
e^{(1)}_{3} &
e^{(2)}_{3} \\
\dd{v_1 x t_\alpha -v_2\over N_1}&
\dd{v_1 x+t_\alpha v_2\over N_2}&
e^{(1)}_{4} &
e^{(2)}_{4}
\end{array}
\right) \, ,
\ee
where
$$
e^{(1)}=v_{L} v_{R}\left(\aac,\abc,\acc,-{s_\beta \over N_4} \right)^T \nn $$
$$
e^{(2)}=v_{L} v_{R}\left(\aad,\abd,\acd,{c_\beta \over N_4} \right)^T.$$

In the above equation, we should understand $s_{\alpha(\beta)}
\equiv \sin\alpha_\pm (\beta_\pm)$, $c_{\alpha(\beta)} \equiv
\cos\alpha_\pm (\beta_\pm)$, $t_{\alpha(\beta)} \equiv \tan
\alpha_\pm (\beta_\pm)$ and, with the same conventions:
\bea
a&=&{v_1 \, v_2 (v_L^2+v_R^2)\over v_R^2 v_L^2 +v_1^2 v_R^2+ v_2^2 v_L^2}
\, , \\
N_1^2&=&v_L^2+u^2+x^2(v_R^2+u^2)t_\alpha^2-4v_1 v_2 x t_\alpha
\, , \\
N_2^2&=&(v_L^2+u^2)t_\alpha^2+x^2(v_R^2+u^2)+4v_1 v_2 x t_\alpha
\, , \\
N_3^2&=&v_R^2 v_L^2 +v_1^2 v_R^2+ v_2^2 v_L^2
\, , \\
N_4^2&=&N_3^2\left(\dd{v_L^2 v_R^2+v_1^2 v_L^2+ v_2^2 v_R^2\over v_L^2
v_R^2+v_2^2 v_L^2+ v_1^2 v_R^2}-a^2\right) \, .
\eea

The explicit form of the orthogonal $4 \times 4$ matrix $C$,
connecting mass and interaction eigenstates in the neutral
CP-odd Higgs sector and defined by eq.~(\ref{rotneu}), is
\be
C=
\left(
\begin{array}{cccc}
\dd{(x^2+y^2) v_R\over M_1}t_\alpha&
\dd{(x^2+y^2) v_R\over M_2}&
d^{(1)}_{1}&
d^{(2)}_{1}\\
\dd{s_W y^2 t_\alpha + x y\over s_W M_1} v_L&
\dd{s_W y^2- x y t_\alpha\over s_W M_2} v_L&
d^{(1)}_{2}&
d^{(2)}_{2}\\
\dd{-s_W x^2 t_\alpha + x y\over s_W M_1} v_1&
\dd{-s_W x^2- x y t_\alpha\over s_W M_2} v_1&
d^{(1)}_{3}&
d^{(2)}_{3}\\
\dd{s_W x^2 t_\alpha - x y\over s_W M_1} v_2&
\dd{s_W x^2+ x y t_\alpha\over s_W M_2} v_2&
d^{(1)}_{4}&
d^{(2)}_{4}
\end{array}
\right) \, ,
\ee
where
$$
d^{(1)}=v_L v_R \left(
\left[{v_1 c_\beta\over v_R M_3}-
{(v_1 b-v_2)s_\beta\over v_R M_4}\right],
\left[{-v_1 c_\beta\over v_L M_3}+
{(v_1 b-v_2)s_\beta\over v_L M_4}\right],
\left[{c_\beta\over M_3}-
{b s_\beta\over M_4}\right],
-{s_\beta\over M_4} \right)^T
\nn $$
$$
d^{(2)}=v_L v_R \left(
\left[{v_1 s_\beta\over v_R M_3}+
{(v_1 b-v_2)c_\beta\over v_R M_4}\right],
\left[{-v_1 s_\beta\over v_L M_3}-
{(v_1 b-v_2)c_\beta\over v_L M_4}\right],
\left[{s_\beta\over M_3}+
{b c_\beta\over M_4}\right],
{c_\beta\over M_4} \right)^T
$$
In the above equation, we should understand $s_{\alpha(\beta)} 
\equiv \sin\alpha_0 (\beta_0)$, $c_{\alpha(\beta)} \equiv 
\cos\alpha_0 (\beta_0)$, $t_{\alpha(\beta)} \equiv \tan
\alpha_0 (\beta_0)$ and, with the same conventions:
\bea
b&=&\dd{v_1 v_2 (v_L^2+v_R^2)\over v_L^2 v_R^2 + v_1^2 (v_L^2+v_R^2)}
\, , \\
M_1^2&=&(v_L^2+u^2)(x^2+y^2+x^2 y^2)+(v_R^2 (x^2+y^2)^2+u^2 x^4+
v_L^2 y^4) t_\alpha^2
\nn \\
& &-2(x^2 u^2-y^2v_L^2) t_\alpha \sqrt{x^2+y^2+x^2 y^2}
\, , \\
M_2^2&=&(v_L^2+u^2)(x^2+y^2+x^2 y^2) t_\alpha^2 +(v_R^2 
(x^2+y^2)^2+u^2 x^4+v_L^2 y^4)
\nn  \\
& &+2(x^2 u^2-y^2v_L^2) t_\alpha \sqrt{x^2+y^2+x^2 y^2}
\, , \\
M_3^2&=&v_L^2 v_R^2 +v_1^2 (v_R^2+v_L^2)
\, , \\
M_4^2&=&M_3^2\left(\dd{v_L^2 v_R^2 +v_2^2 (v_R^2+v_L^2)
\over v_L^2 v_R^2 +v_1^2 (v_R^2+v_L^2)}-b^2\right) \, .
\eea

In terms of the mass eigenstates in the gauge boson sector,
$(W,W')$ and $(A,Z,Z')$, the trilinear gauge boson vertices 
read:
\bea
+ i g \left(
c_{W} c_{0} c_{\pm}^{2} +{x^{2} \over \sqrt{x^{2}+y^{2}}} s_{0}
s_{\pm}^{2}
 - {x^{2} y^{2} \over
\sqrt{x^2 + y^2 } \sqrt{x^2 + y^2 + x^2 y^2}} c_{0} s_{\pm}^{2}
\right)
\times \nn \\
 ( W_{\nu}^{-} \partial_{\mu} W_{\nu}^{+} Z_{\mu}
+ W_{\mu}^{+} \partial_{\mu} W_{\nu}^{-} Z_{\nu} +
 W_{\mu}^{-} W_{\nu}^{+} \partial_{\mu} Z_{\nu} ) \nn \\
+ i g \left(
c_{W} c_{0} s_{\pm}^{2} +{x^{2} \over \sqrt{x^{2}+y^{2}}} s_{0}
c_{\pm}^{2}
 - {x^{2} y^{2} \over
\sqrt{x^2 + y^2 } \sqrt{x^2 + y^2 + x^2 y^2}} c_{0} c_{\pm}^{2}
\right) \times \nn \\
 ( W_{\nu}^{\prime -} \partial_{\mu} W_{\nu}^{\prime +} Z_{\mu}
+ W_{\mu}^{\prime +} \partial_{\mu} W_{\nu}^{\prime -} Z_{\nu} +
 W_{\mu}^{\prime -} W_{\nu}^{\prime +} \partial_{\mu} Z_{\nu} ) \nn
\\
+ i g \left(
- c_{W} s_{0} c_{\pm}^{2} +{x^{2} \over \sqrt{x^{2}+y^{2}}} c_{0}
s_{\pm}^{2}
 + {x^{2} y^{2} \over
\sqrt{x^2 + y^2 } \sqrt{x^2 + y^2 + x^2 y^2}} s_{0} s_{\pm}^{2}
\right)
\times \nn \\ ( W_{\nu}^{-} \partial_{\mu} W_{\nu}^{+} Z_{
\mu}^{\prime}
+ W_{\mu}^{+} \partial_{\mu} W_{\nu}^{-} Z_{\nu}^{\prime} +
 W_{\mu}^{-} W_{\nu}^{+} \partial_{\mu} Z_{\nu}^{\prime} ) \nn \\
- i g c_{\pm} s_{\pm} \left( c_{W} c_{0} - {x^{2} \over \sqrt{x^{2} +
y^{2}} } s_{0} +
{x^{2} y^{2} \over
\sqrt{x^2 + y^2 } \sqrt{x^2 + y^2 + x^2 y^2}} c_{0}\right)
\times \nn \\
 ( W_{\nu}^{\prime -} \partial_{\mu} W_{\nu}^{+} Z_{\mu}
+ W_{\mu}^{\prime +} \partial_{\mu} W_{\nu}^{-} Z_{\nu} +
 W_{\mu}^{\prime -} W_{\nu}^{+} \partial_{\mu} Z_{\nu}  + \nn \\
  W_{\nu}^{-} \partial_{\mu} W_{\nu}^{\prime +} Z_{\mu}
+ W_{\mu}^{ +} \partial_{\mu} W_{\nu}^{\prime -} Z_{\nu} +
 W_{\nu}^{\prime +} W_{\mu}^{-} \partial_{\mu} Z_{\nu} ) \nn \\
- i g c_{\pm} s_{\pm} \left(- c_{W} s_{0} - {x^{2} \over \sqrt{x^{2}
+
y^{2}} } c_{0} -
{x^{2} y^{2} \over
\sqrt{x^2 + y^2 } \sqrt{x^2 + y^2 + x^2 y^2}} s_{0}\right) \times \nn
\\
 ( W_{\nu}^{\prime -} \partial_{\mu} W_{\nu}^{+} Z_{\mu}^{ \prime}
+ W_{\mu}^{\prime +} \partial_{\mu} W_{\nu}^{-} Z_{\nu}^ {\prime} +
 W_{\mu}^{\prime -} W_{\nu}^{+} \partial_{\mu} Z_{\nu}^{ \prime}  +
\nn
\\
  W_{\nu}^{-} \partial_{\mu} W_{\nu}^{\prime +} Z_{\mu}^{ \prime}
+ W_{\mu}^{ +} \partial_{\mu} W_{\nu}^{\prime -} Z_{\nu}^{ \prime} +
 W_{\nu}^{\prime +} W_{\mu}^{-} \partial_{\mu} Z_{\nu}^{ \prime} )
\nn
\\
+ i g \left(
- c_{W} s_{0} s_{\pm}^{2} +{x^{2} \over \sqrt{x^{2}+y^{2}}} c_{0}
c_{\pm}^{2}
 + {x^{2} y^{2} \over
\sqrt{x^2 + y^2 } \sqrt{x^2 + y^2 + x^2 y^2}} s_{0} c_{\pm}^{2}
\right)
\times \nn \\ ( W_{\nu}^{\prime -} \partial_{\mu} W_{\nu}^{\prime +}
Z_{
\mu}^{\prime}
+ W_{\mu}^{\prime +} \partial_{\mu} W_{\nu}^{\prime -}
Z_{\nu}^{\prime}
+
 W_{\mu}^{\prime -} W_{\nu}^{\prime +} \partial_{\mu}
Z_{\nu}^{\prime} )
\nn \\
+i g s_{W} \left( A_{\mu} W_{\nu}^{-}\partial_{\mu} W_{\nu}^{+}+
 A_{\nu} W_{\mu}^{+}\partial_{\mu} W_{\nu}^{-}+
W_{\mu}^{-} W_{\nu}^{+} \partial_{\mu} A_{\nu} \right)
\nn \\
+i g s_{W} \left( A_{\mu} W_{\nu}^{\prime -}\partial_{\mu}
W_{\nu}^{\prime +}+
 A_{\nu} W_{\mu}^{\prime +}\partial_{\mu} W_{\nu}^{\prime -}+
W_{\mu}^{\prime -} W_{\nu}^{\prime +} \partial_{\mu} A_{\nu} \right)
\, .
\eea
It is understood that we should add to the previous couplings
their hermitean conjugates, and we have used the conventions
$s_{\pm} \equiv \sin \alpha_{\pm}$, $c_{\pm} \equiv \cos 
\alpha_{\pm}$, $s_0 \equiv \sin \alpha_0$, $c_0 \equiv \cos 
\alpha_0$, $s_W \equiv \sin \theta_W$, $c_W \equiv \cos \theta_W$.

The cubic bosonic couplings involving the SM-like Higgs $h$, defined
by eq.~(\ref{smhiggs}), and the gauge boson mass eigenstates are,
in the same conventions as before:
\bea
-&{1 \over \sqrt{v_{L}^{2}+v_{1}^{2}+v_{2}^{2}}}& {h \over {2}}
g_{R}^{2} v_{R}^{2} s_{\pm} c_{\pm} ( W_{\mu}^{-} W^{\prime \mu +} +
W_{\mu}^{+} W^{\prime \mu -}) \nn \\
+&{1 \over \sqrt{v_{L}^{2}+v_{1}^{2}+v_{2}^{2}}}& {h}
(2 m_{W}^{2} - \dd\frac{1}{2}g_{R}^{2} v_{R}^{2} s_{\pm}^{2}) 
W_{\mu}^{-} W^{\mu +}
\nn
\\ +&{1 \over \sqrt{v_{L}^{2}+v_{1}^{2}+v_{2}^{2}}}& {h}
(2 m_{W}^{\prime 2} - \dd\frac{1}{2}g_{R}^{2} v_{R}^{2} c_{\pm}^{2})
{W^{\prime}}_{\mu}^{-}
{W^{\prime}}^{\mu +} \nn \\
- &{1 \over \sqrt{v_{L}^{2}+v_{1}^{2}+v_{2}^{2}}}& {h\over 2}
(g_{R}^{2} + {\tilde{g}}^{2}) v_{R}^{2} s_{0} c_{0}  Z_{\mu}
Z^{\prime
\mu}
 \nn \\
+&{1 \over \sqrt{v_{L}^{2}+v_{1}^{2}+v_{2}^{2}}}& {h \over 2}
(2 m_{Z}^{2} - \dd\frac{1}{2}
(g_{R}^{2} + {\tilde{g}}^{2}) v_{R}^{2} s_{0}^{2})
Z_{\mu} Z^{\mu} \nn \\
+&{1 \over \sqrt{v_{L}^{2}+v_{1}^{2}+v_{2}^{2}}}& {h \over 2}
(2 m_{Z}^{\prime 2} - \dd\frac{1}{2}
(g_{R}^{2} + {\tilde{g}}^{2}) v_{R}^{2}
c_{0}^{2})
{Z^{\prime}}_{\mu} {Z^{\prime}}^{\mu} \, .
\eea

\newpage
\appendixB{Appendix~B}

We collect in this appendix the explicit expressions, 
valid at the classical level, for the most important partial
decay rates of the $W'$ and $Z'$ bosons in the FP model.

From the explicit expressions of the charged currents,
given in section 2.2.1, we can easily derive the vector and 
axial couplings of fermion doublets $f \equiv \{ f_1,f_2 
\}$ to the charged vector boson $W'$:
\be
v_f(W') = 
a_f(W') = 
- s_\pm a_f(W_L)  {g \over \sqrt{2}} T_L (f_L) \, .
\ee
The partial decay rates of $W'$ into fermion pairs are then 
given by the standard formulae:
\be
\Gamma(W' \rightarrow f_1 \overline{f_2} ) =
C_f |K_{f_1 f_2}|^2
{m_{W'} \over 12 \pi} [1 - 2 (x'_1+x'_2)+(x'_1-x'_2)^2]^{1/2}
\nn
\ee
\be
\left\{ \left[ v_f^2(W') + a_f^2(W') \right] \left[
1-{x'_1 +x'_2 \over 2}- {(x'_1-x'_2)^2 \over 2} \right]
+ 3 \left[ v_f^2(W') - a_f^2(W') \right] (x'_1
x'_2)^{1/2} \right\} \, ,
\ee
where $C_f=1,3$ is the colour factor, $x'_{1,2} \equiv 
m_{f_{1,2}}^2/m_{W'}^2$ and $K$ is the CKM matrix.

From the explicit expressions of the neutral currents,
given in section 2.2.1, we can easily derive the vector 
and axial couplings of fermions to the neutral vector
boson $Z'$:
\be
\begin{array}{rcl}
v_f(Z_L) & = & {e \over s_W c_W} \left[
{T_{3L}(f_L) \over 2} - Q(f) s_W^2 \right]
\, , \\
a_f(Z_L) & = & {e \over s_W c_W} {T_{3L}(f_L) 
\over 2}
\, , \\
v_f(Z_R) & = & - {e \over 2 s_W \sqrt{x^2+y^2}} \left\{
y^2 [ \tilde{Y}(f_L) + \tilde{Y}(f_R) ] \right\}
\, , \\
a_f(Z_R) & = & - {e \over 2 s_W \sqrt{x^2+y^2}} \left\{
y^2 [ \tilde{Y}(f_L) - \tilde{Y}(f_R) ] \right\}  \, .
\end{array}
\ee
\be
v_f(Z') = - s_0 v_f(Z_L) + c_0 v_f(Z_R) \, ,
\;\;\;\;\;
a_f(Z') = - s_0 v_f(Z_L) + c_0 v_f(Z_R) \, .
\ee
The $Z'$ partial decay rates into fermion pairs are 
then given by the standard formulae:
\be
\Gamma(Z' \rightarrow f \overline{f} ) =
{m_{Z'} \over 12 \pi} (1 - 4 \eta_f')^{1/2}
C_f \left\{ v_f^2(Z') + a_f^2(Z') + 2 [ v_f^2(Z') -
2 a_f^2(Z') ] \eta_f' \right\} \, ,
\ee
where $C_f=1,3$ is the colour factor and 
$\eta_f' \equiv m_f^2/m_{Z'}^2$.

From the explicit expressions of the trilinear gauge boson
vertices, given in appendix~A, we can easily derive the 
$Z'$ and $W'$ decay rates into gauge bosons, when the
processes are kinematically allowed:
\be
\Gamma(Z' \rightarrow W^+ W^- ) =
{m_{Z'} \over 192 \pi} \eta_{WZ'}^{-2}
(1 - 4 \eta_{WZ'})^{3/2}
(1 + 20 \eta_{WZ'} + 12 \eta_{WZ'}^2)
\delta^2(Z'WW) \, ,
\ee
where $\eta_{WZ'} \equiv m_W^2 / m_{Z'}^2$ and
\be
\delta (Z'WW) = g \left( c_W s_0 c_{\pm}^2 -
{x \over y} t_W  c_0 s_{\pm}^2
- s_W t_W s_0 s_{\pm}^2 \right) \, .
\ee
\be
\Gamma(Z' \rightarrow W W') =
{m_{Z'} \over 192 \pi}  \eta_{WZ'}^{-1} \eta_{W'Z'}^{-1}
\left[ \left( 1 - \eta_{WZ'} + \eta_{W'Z'} \right)^2
- 4 \eta_{W'Z'} \right]^{3/2}
\ee
\be
\left[ 1 +  10 \left(  \eta_{W'Z'} + \eta_{WZ'} \right)
+ \eta_{W'Z'}^2 + \eta_{WZ'}^2 + 10 \eta_{W'Z'} \eta_{WZ'}
\right]  \delta^2(Z'WW') \, ,
\ee
where $\eta_{W'Z'} \equiv m_{W'}^2 / m_{Z'}^2$, $\eta_{WZ'}
\equiv m_W^2 / m_{Z'}^2$, and
\be
\delta (Z'WW') =  g s_{\pm} c_{\pm} 
\left( {s_0 \over c_W} +
{x \over y} t_W c_0 \right)  \, .
\ee
\be
\Gamma(Z' \rightarrow W' W') =
{m_{Z'} \over 192 \pi} \eta_{W'Z'}^{-2}
(1 - 4 \eta_{W'Z'})^{3/2}
(1 + 20 \eta_{W'Z'} + 12 \eta_{W'Z'}^2)
\delta^2(Z'W'W') \, ,
\ee
where $\eta_{W'Z'} \equiv m_{W'}^2 / m_{Z'}^2$ and
\be
\delta (Z'W'W') = g \left( - c_W s_0 s_{\pm}^2 +
{x \over y} t_W c_0 c_{\pm}^2
+ s_W t_W s_0 c_{\pm}^2 \right) \, .
\ee
\be
\Gamma(W' \rightarrow W Z) =
{m_{W'} \over 192 \pi}  \eta_{ZW'}^{-1} \eta_{WW'}^{-1}
\left[ \left( 1 - \eta_{ZW'} + \eta_{WW'} \right)^2
- 4 \eta_{WW'} \right]^{3/2}
\ee
\be
\left[ 1 +  10 \left(  \eta_{WW'} + \eta_{ZW'} \right)
+ \eta_{WW'}^2 + \eta_{ZW'}^2 + 10 \eta_{WW'} \eta_{ZW'}
\right]  \delta^2(W'WZ) \, ,
\ee
where $\eta_{WW'} \equiv m_W^2 / m_{W'}^2$, $\eta_{ZW'}
\equiv m_Z^2 / m_{W'}^2$, and
\be
\delta (W'WZ) = g c_{\pm} s_{\pm} \left( {c_0 \over c_W}
- s_0 {x \over y} t_W \right) \, .
\ee

From the explicit expressions of the cubic bosonic couplings 
involving the SM-like Higgs $h$ and pairs of vector bosons,
given in appendix~A, we can easily derive the $Z'$ and $W'$ 
decay rates into final states involving gauge and Higgs 
bosons, when the processes are kinematically allowed:
\be
\Gamma(Z' \rightarrow Z h) =
{m_{Z'} \over 192 \pi}
\left[ \left( 1 - \eta_{hZ'} + \eta_{ZZ'} \right)^2
- 4 \eta_{ZZ'} \right]^{1/2}
\ee
\be
\left[ 1 +  \eta_{hZ'}^2 + \eta_{ZZ'}^2 +
10 \eta_{ZZ'} - 2 \eta_{hZ'} - 2 \eta_{hZ'}
\eta_{ZZ'} \right]
\left[ { \delta(Z'Zh) \over m_Z } \right]^2 \, ,
\ee
\be
\Gamma(W' \rightarrow W h) =
{m_{W'} \over 192 \pi}
\left[ \left( 1 - \eta_{hW'} + \eta_{WW'} \right)^2
- 4 \eta_{WW'} \right]^{1/2}
\ee
\be
\left[ 1 +  \eta_{hW'}^2 + \eta_{WW'}^2 +
10 \eta_{WW'} - 2 \eta_{hW'} - 2 \eta_{hW'}
\eta_{WW'} \right]
\left[ { \delta(W'Wh) \over m_W } \right]^2 \, ,
\ee
where $\eta_{hZ'} \equiv m_h^2 / m_{Z'}^2$, $\eta_{hW'}
\equiv m_h^2 / m_{W'}^2$, and
\be
\delta (Z'Zh) =
{ (g_R^2 + \tilde{g}^2 ) v_R^2
\over
2 \sqrt{v_L^2 + v_1^2 + v_2^2}}
s_0 c_0 \, ,
\ee
\be
\delta (W'Wh) =
{ g_R^2  v_R^2
\over
2 \sqrt{v_L^2 + v_1^2 + v_2^2}}
s_{\pm} c_{\pm} \, .
\ee

It is important to notice the following asymptotic relations,
valid when $m_{W',Z'} \gg m_{W,Z}$:
\be
\delta^2(Z'WW) \rightarrow g^2 c_W^2 s_0^2 \, ,
\;\;\;\;\;
\delta^2(W'WZ) \rightarrow {g^2 s_{\pm}^2 \over c_W^2} \, ,
\ee
\be
\delta^2(Z'Zh) \rightarrow {g^2 s_0^2 m_{Z'}^4 \over m_W^2} \, ,
\;\;\;\;\;
\delta^2(W'Wh) \rightarrow {g^2 s_{\pm}^2 m_{W'}^4 \over m_W^2} \,
,
\ee
\be
{\Gamma(Z' \rightarrow W W) \over \Gamma(Z' \rightarrow Z h)}
\rightarrow 1 \, ,
\;\;\;\;\;
{\Gamma(W' \rightarrow W Z) \over \Gamma(W' \rightarrow W h)}
\rightarrow 1 \, .
\ee

For convenience, we rewrite the expressions for the triple
bosonic couplings at the lowest order in $\alpha_{\pm},
\alpha_0$:
\be
\delta (Z'WW) = g c_W \alpha_0  \, ,
\;\;\;\;\;
\delta (W'WZ) = {g \over c_W} \alpha_{\pm} \, ,
\ee
\be
\delta (Z'WW') = g {x \over y} t_W \alpha_{\pm}  \, ,
\;\;\;\;\;
\delta (Z'W'W') = g {x \over y} t_W \, ,
\ee
\be
\delta (Z'Zh) =
{ (g_R^2 + \tilde{g}^2 ) v_R^2
\over
2 \sqrt{v_L^2 + v_1^2 + v_2^2}}
\alpha_0 \, ,
\;\;\;\;\;
\delta (W'Wh) =
{ g_R^2  v_R^2
\over
2 \sqrt{v_L^2 + v_1^2 + v_2^2}}
\alpha_{\pm} \, .
\ee
\newpage
\end{document}